\documentclass[12pt]{article}

\makeatletter \@addtoreset{equation}{section} \makeatother
\makeatletter \@addtoreset{figure}{section} \makeatother

\def\CA{{\cal A}}
\def\CF{{\cal F}}
\def\CJ{{\cal J}}
\def\CL{{\cal L}}
\def\CN{{\cal N}}
\def\CR{{\cal R}}
\def\CT{{\cal T}}

\def\bA{{\mathbf A}}\def\bB{{\mathbf B}}\def\bE{{\mathbf E}}
\def\NA{{\mathbf \nabla}}

\def\tr{{\rm tr}}

\newcommand{\be}{\begin{equation}}
\newcommand{\ee}{\end{equation}}
\newcommand{\bea}{\begin{eqnarray}}
\newcommand{\eea}{\end{eqnarray}}
\newcommand{\ba}{\begin{array}}

\def\ba{{\bar a}}
\def\be{{\bar e}}

\begin{document}
\begin{titlepage}
\vfill
\begin{flushright}
{\tt\normalsize KIAS-P11082}\\

\end{flushright}
\vfill
\begin{center}
{\Large\bf D-Brane Anomaly Inflow Revisited}

\vskip 1cm

Heeyeon Kim\footnote{\tt hykim@phya.snu.ac.kr}$^\dagger$
and Piljin Yi\footnote{\tt piljin@kias.re.kr}$^\ddagger$

\vskip 5mm
$^\dagger${\it Department of Physics and Astronomy, Seoul
National University, \\Seoul 151-147, Korea}
\vskip 3mm $^\ddagger${\it School of Physics, Korea Institute
for Advanced Study, Seoul 130-722, Korea}

\end{center}
\vfill

\begin{abstract}
\noindent
Axial and gravitational anomaly of field theories,
when embedded in string theory, must be accompanied by canceling
inflow. We give a self-contained overview for various world-volume
theories, and clarify the role of smeared magnetic
sources in I-brane/D-brane cases. The proper anomaly descent of
the source, as demanded by regularity of RR field strengths $H$'s,
turns out to be an essential ingredient. We show how this allows
correct inflow to be generated for all such theories, including
self-dual cases, and also that the mechanism is now insensitive
to the choice between the two related but inequivalent forms of
D-brane Chern-Simons couplings. In particular, $SO(6)_R$ axial anomaly
of $d=4$ maximal SYM is canceled by the inflow onto D3-branes
via the standard minimal coupling to $C_4$. We also propose how,
for the anomaly cancelation, the four types of Orientifold planes
should be coupled to the spacetime curvatures, of which conflicting
claims existed previously.

\end{abstract}

\vfill
\end{titlepage}

\parskip 0.1 cm
\tableofcontents\newpage
\renewcommand{\thefootnote}{\#\arabic{footnote}}
\setcounter{footnote}{0}

\parskip 0.2 cm

\section{Introduction}

Field theories with extended supersymmetries are equipped with
R-symmetry and sometimes other accidental global symmetries, which
become typically anomalous at one-loop. In a slight abuse of
nomenclature, we will call them collectively the axial anomaly.
While the axial anomaly is not a consistency issue at the level
of field theories, it becomes one when one realizes such a
theory as a part of string theory or M-theory. In the latter
context, the R-symmetry and the global symmetries are
realized as a part of ten- or eleven-dimensional diffeomorphism
invariance, whose anomaly will render the system gravitationally
inconsistent, so there has to be a canceling contribution from
the underlying string theory or M-theory. Clearly the gravitational
anomaly belongs to the same class, so must be considered
simultaneously with the axial anomaly.
For many field theories arising as world-volume dynamics of
M-branes and D-branes, the anomaly inflow has been cataloged
and shown to cancel the one-loop anomaly of the field theory.

The anomaly inflow originates from the topological couplings and
resulting modification of the Bianchi identities of
anti-symmetric tensor fields. A prime example of this is
the Green-Schwarz mechanism for type I and Heterotic
string theory. Other than this, there are two principal systems
where the anomaly inflow is important. One class is the M5-brane,
and the others are the D-branes and I-branes (meaning intersection
of D-branes). For the M5-branes, inflow occurs from
a spacetime topological coupling to a lower dimensional
world-volume, and can be understood more easily as a direct
result of the modified Bianchi identity
\cite{Duff:1995wd}\cite{Witten}\cite{FHMM}. As far as anomaly
inflow goes, M5-brane represents the best understood example,
although its world-volume field theory, namely $(2,0)$ theory,
and in particular how the one-loop anomaly arises remain largely
mysterious.

On the other hand,
the easiest examples of world-volume field theories with one-loop
axial/gravitational anomaly are maximally supersymmetric
Yang-Mills theories in $d=4,6,8,10$ dimensions, some of which can be
realized on coincident D-branes. As we review in section 2, anomalies
in $d$ dimensions are dictated by topological $(d+2)$-form polynomials,
which, for the maximally supersymmetric Yang-Mills theories, are\footnote{
For $d=4k$,  one actually has $-2ch_{S_-}(F_R)$ in place of
 $[ch_{S_+}(F_R)-ch_{S_-}(F_R)] $, but  this substitution
 does not affect the  relevant $(4k+2)$-form part.}
\begin{eqnarray}\label{agw}
(-1)^{d/2}\pi \cdot\left( [ch_{\rm adj}(\CF)+l]\wedge
 \CA(R)\wedge \left[ch_{S^+}(F_R)-ch_{S^-}(F_R)\right]\right)\biggl\vert_{(d+2)-form},
\end{eqnarray}
with the Chern class $ch$, the A-roof genus $\CA$, gauge field strength
$\CF$, spacetime curvature tensor $R$, and the number of $U(1)$
factors $l$ in the gauge group \cite{AlvarezGaume:1983ig}.
$F_R$ is the field strength of the
external R-gauge field and $S_\pm$ are the chiral and
anti-chiral spinor representations of the R-symmetry
$SO(10-d)_R$. We are using a Lorentzian signature
$(-++\cdots +)$.\footnote{Here we took the computations of Ref. [6]
in the Euclidean signature $(+ + + \cdots +)$, and weak-
rotated it to $ (- + +\cdots +)$. This produces the overall sign in (1.1).}

With $G=U(n), \, SO(n), \, Sp(n)$, this is a world-volume theory of D$p$-branes
with or without various Orientifold planes. As such, the gravitational curvature $R$
and the R-symmetry curvature $F_R$ are, respectively, associated with the
tangent bundle $\CT$ and the normal bundle $\CN$ of the world-volume
and we will henceforth rewrite the anomaly polynomial as
\begin{equation}\label{one-loop}
 I^{1-loop}_{d=p+1}=(-1)^{(p+1)/2}\pi \cdot \left([ch_{\rm adj}(\CF)+l]\wedge
 \CA(\CT)\wedge [ch_{S^+}(\CN)-ch_{S^-}(\CN)]\right)\biggl\vert_{(d+2)-form},
\end{equation}
where $l=1$ for $U(n)$ and $0$ otherwise.
As such, the anomaly must be canceled by other stringy contributions,
since the diffeomorphism invariance must be preserved. This anomaly
is null for $d=2$, so we expect a canceling inflow for $4\le d\le 10$.

For cancelation of such anomaly, one must understand inflow onto
D-branes, which is a little more involved than the M5 case. One reason is
that the relevant topological coupling (with $2\pi\sqrt{\alpha'}=1$), say, for each
stacks of coincident D$p$-branes with charge $\mu_p$,
\begin{equation}\label{CS2}
S_{CS}=\frac{\mu_p}{2} \int_{Dp} \;\sum_{r\le p} s^*(C_{r+1})\wedge
ch(\CF)\wedge \sqrt{\frac{\CA(\CT)}{\CA({\CN})}}\ ,
\end{equation}
lives in the same world-volume as the relevant one-loop
anomaly. Throughout this paper, we  denote the pull-back of spacetime
forms to the relevant world-volume by $s^*$. The overall factor $1/2$
\cite{CY} may be a little puzzling, but  is a consequence of
the so-called duality symmetric formulation
of $C$'s. The latter is
necessary because electromagnetic dual pairs of RR fields always
act together in I-brane/D-brane inflow mechanism. Most importantly,
this factor $1/2$ disappears in the field equation derived
from the duality symmetric formulation, which resolves
all the potential conflicts and, in particular, helps one to
recover the Dirac quantization conditions with the usual,
properly quantized charges $\mu_p$.\footnote{See Appendices A and B.}

A somewhat unexpected result in literatures, though,
is that, under the standard procedure, one finds the right canceling inflow only if the following alternative
and inequivalent form of these couplings is used  \cite{CY}\cite{GHM},
\begin{equation}\label{CS0'}
\frac{\mu_p}{2}\int_{Dp}  \;\left( N_p\,s^*(C_{p+1})
\pm \sum_{r<p} s^*(H_{r+2})\wedge(\cdots)\right),
\end{equation}
where $H$ is the gauge-invariant field strengths of $C$ and
$N_p$ is the number of coincident D$p$-branes. The ellipsis
represents the odd-form Chern-Simons densities from
$ch(\CF){\CA(\CT)}^{1/2}\CA({\CN})^{-1/2}$. See section 2.3
and equation (\ref{CS3}) for complete details.
The two would be equivalent if $H= dC$, but this is not
the case because $dH\neq 0$ in general. This failure
of the Bianchi identity is the main mechanism that underlies
the inflow, and one traditionally finds two different inflow from these
two sets of couplings.
Furthermore, the case of D3-branes proved to be
fairly subtle among these examples. The one-loop anomaly polynomial (\ref{agw})
reduces for $d=4$ to
\begin{equation}
 {\rm dim}\, G\times\frac{1}{24\pi^2}\, {\rm tr}_{S^+} F^3_R~,
\end{equation}
and is purely axial with $SO(6)_R$ R-symmetry.
${\rm dim\,} G$ is the dimension of the gauge group $G$, and
other terms cancel out thanks to the reality of the adjoint
representation.
Yet, the conventional procedure involving the above
topological couplings on D-branes fails to generate any
inflow at all for D3-branes which, in view of how various D-branes
are connected to each other by T-dualities, sounds quite odd.

In this note, we wish to revisit these anomaly inflow and clarify
some of these finer points. We will
emphasize on how we must regularize the Bianchi identity and
magnetic sources. The regularization is not necessary for
the simplest type of inflow, such as gravitational anomaly
of M5-brane theory. For others, regularization is essential.
For the axial anomaly of M5-branes, this has been exploited
carefully by Freed, Harvey, Minasian, and Moore (FHMM) \cite{FHMM}, while
it also played some role in the Cheung-Yin's  (CY) \cite{CY} elaboration of
I-brane inflow arguments by Green, Harvey, and Moore (GHM) \cite{GHM}.
At the end of day, however, several unsatisfactory aspects
remain, one of which is the apparent absence of D3 anomaly
inflow we already mentioned. In this note, we combine the ideas
of FHMM and of GHM/CY to address the D-brane and I-brane anomaly
inflow again and resolve such outstanding issues.

In section 2 and 3, we review various  anomaly inflow
mechanisms in string theory and in M-theory. In section 2, after
a brief review of the consistent anomaly and the simplest inflow
mechanism (gravitational anomaly on a M5-brane),
we retrace the steps taken by CY for D-brane and I-brane
inflow. In particular, we note that, to obtain the desired
inflow, they had to use the
modified Chern-Simons couplings (\ref{CS0'}) rather than
the more natural looking one (\ref{CS2}) \cite{CY}.
We will  delineate how the usual procedure also fails to
produce necessary D3-brane anomaly inflow.
In section 3, we turn to the anomaly inflow onto M5-branes associated
with the $SO(5)_R$ symmetry by FHMM. Although the mechanism
of inflow here is qualitatively different from other examples, we
will learn an important lesson that should be
applied to the D-brane and I-brane story.

In section 4, we reconsider the D-brane and I-brane inflow
by requiring both source terms in the Bianchi identity
and the RR field strengths to be regular, which are of
course interconnected to each other. This requirement
modifies the solution to the Bianchi identity, in a manner that
fundamentally changes gauge transformation properties of the
RR gauge fields. With this revised transformation rule,
we re-derive the anomaly inflow for D-branes and I-branes,
and find that the standard Chern-Simons coupling of type
(\ref{CS2}) generates all the necessary anomaly inflow,
as well as (\ref{CS0'}).
In particular, this includes ``self-dual" cases like
the D3-branes. Despite the naive
difficulties with these ``self-dual" cases, the correct
inflow arises without a special treatment.

In section 5, we extend all these discussion to systems
involving Orientifold planes. In literature, there
appears to be partially conflicting claims regarding what should be
the right (gravitational) Chern-Simons couplings on the four
types of Orientifold planes
\cite{DFM}\cite{Scrucca:1999jq}\cite{Mukhi}\cite{Henry}\cite{JFO}.
Here we settle this
by requiring cancelation of the axial and gravitational anomaly of
orthogonal and symplectic gauge theories  and also demanding that
the inflow to be canceled by
closed string one-loop contribution is independent of the
Orientifold type.

\section{Anomaly Inflows}\label{sec2}

\subsection{Consistent Anomaly }

Recall  \cite{anomaly} that the so-called consistent anomaly on $d$ dimensions is
represented by a characteristic polynomial of rank $d+2$,
say $X(F,R,\cdots)$, of curvature tensors, via a descent relation,
\begin{equation}
X_{d+2}= dX^{(0)}_{d+1}\ , \qquad \delta X_{d+1}^{(0)}=dX_{d}^{(1)}\ ,
\end{equation}
such that the anomaly associated with $X_{d+2}$ is actually an
integral of $X_{d}^{(1)}$. Note that this procedure is
ambiguous since $X_{d+1}^{(0)}\rightarrow X_{d+1}^{(0)}+dZ_d$
with $d$-form $Z_d$. However, this is not an issue because
the additional anomaly due to this shift is  $\delta Z_d$
and thus cancelable by a local counter-term $-Z_d$.
This simple observation gives us a useful generality about
anomaly: when $X=Y\wedge \tilde Y$, the anomaly
due to $X$ can be expressed as
\begin{equation}
\beta Y\wedge \tilde Y^{(1)} +(1-\beta )Y^{(1)}\wedge \tilde Y\ ,
\end{equation}
since a counter term of type $Y^{(0)}\wedge\tilde Y^{(0)}$,
provided that both the factors exist, can always shift the
parameter $\beta$. In this note, we will  mostly use the
symmetric version $\beta=1/2$.

It is important to note that when one of the two factors, say $\tilde Y$, is 0-form and thus
constant, we must use $Y^{(1)}\tilde Y$ and vice versa. More
generally, when $Y$ includes a 0-form constant, say $Y=Y_0+dY^{(0)}$
etc, we have instead
\begin{equation}
(Y\wedge \tilde Y)^{(1)}=Y_0\tilde Y^{(1)}+Y^{(1)}\tilde Y_0
+(dY^{(0)}\wedge \tilde Y^{(1)}+ Y^{(1)}\wedge d\tilde Y^{(0)} )/2 \ ,
\end{equation}
which is different from
$
(Y\wedge \tilde Y^{(1)}+ Y^{(1)}\wedge \tilde Y )/2 \ .
$

In usual field theories, anomaly arises from one-loop.
In Fujikawa's path integral formulation, this can be understood
as a failure of the path integral measure to respect symmetry
of the Lagrangian.
There are also situations where such an
anomaly is present at tree level. The Wess-Zumino-Witten
term  of chiral perturbation theory \cite{WZW}, which captures one-loop flavor anomaly of QCD,
is probably one of the oldest such example. This type of tree-level
anomaly in low energy effective action is there because the `t Hooft
anomaly matching condition must be respected. Another important
class of tree-level anomaly is called the anomaly inflow, in string
theory setting, which arises to cancel would-be harmless anomaly
associated with global symmetries of a field theory because the
global symmetries are typically gauged once embedded in string
theory. Regardless of precise mechanism of how it is generated,
however, anomaly can be cast into the above form as a descent from
a characteristic class of rank $d+2$.

\subsection{M5-Brane Gravitational Anomaly Inflow }

Perhaps the simplest example of such an inflow can be found
in the context of a single M5-brane, whose world-volume theory is
a tensor multiplet theory in six dimensions. To set a
consistent convention, let us write the 11-dimensional
supergravity action as
\begin{eqnarray}
S_{11}=\frac{1}{2\kappa_{11}^2}\left[\int\sqrt{-g}\left(R-\frac12|\,G_4|^2\right)
-\frac16\int C_3\wedge G_4\wedge G_4\right]+\mu_{M2}\int C_3\wedge I_8\ ,
\end{eqnarray}
with the three-form gauge field of M-theory $C_3$, its field strength $G_4$,
M2-brane tension $\mu_{M2}$, and the
8-form polynomial $I_8$ of the spacetime curvature two-form
\begin{equation}
\qquad I_8= -\frac{1}{48}\left(p_2(R)-\frac14p_1(R)^2\right)\ ,
\end{equation}
where $p_n$'s are the Pontryagin classes. See Appendix C for definition of
characteristic classes we will encounter in this  note. In terms of the 11-dimensional
Planck length $l_p$, recall that
\begin{equation}
\frac{1}{2\kappa_{11}^2}=\frac{2\pi}{(2\pi l_p)^9}\ ,\quad \mu_{M2}=\frac{2\pi}{(2\pi l_p)^3}\ ,
\quad \mu_{M5}=\frac{2\pi}{(2\pi l_p)^6}\ .
\end{equation}
For the M5-brane anomaly inflow discussion, we are using the unit
$2\pi l_p=1$, whereby $1/2\kappa_{11}^2=\mu_{M2}=\mu_{M5}=2\pi$.

In the presence of a M5, which couples to
$C_3$ magnetically,
\begin{equation}\label{Bianchi}
dG_4=2\kappa_{11}^2\mu_{M5} \delta_{M5} =\delta_{M5}\ ,
\end{equation}
this coupling induces a tree-level anomaly
on the M5 world-volume. The argument starts with the alternate
form of the topological coupling
\begin{equation}
\mu_{M2}\int C_3\wedge I_8\quad\rightarrow\quad \mu_{M2}\int G_4\wedge I_7^{(0)}\ , \qquad dI_7^{(0)}=I_8\ ,
\end{equation}
which varies under the eleven dimensional diffeomorphism \cite{Duff:1995wd} as
\begin{equation} \label{inflow}
\mu_{M2}\int
G_4\wedge  \delta I_7^{(0)} = -\mu_{M2} \int dG_4\wedge I_6^{(1)}
=  -2\pi \int_{M5}I_6^{(1)} \ ,
\end{equation}
as  $2\kappa_{11}^2\mu_{M2}\mu_{M5} =2\pi $. This inflow
is capable of canceling world-volume
anomaly of the form,
\begin{equation}
2\pi I_8(\CT\oplus\CN)=
-2\pi\times \frac{1}{48}\left(p_2({\cal T})+p_2({\cal N})
-\frac{(p_1({\cal T})-p_1({\cal N}))^2}{4}\right)\ ,
\end{equation}
where $\CT$ and $\CN$ denote tangent and normal bundles of the M5-brane.

On the other hand, the one-loop anomaly polynomial of a single tensor
multiplet is
\begin{equation}
2\pi {\cal J}_8=
-2\pi\times\frac{1}{48}\left(p_2({\cal T})-p_2({\cal N})
-\frac{(p_1({\cal T})-p_1({\cal N}))^2}{4}\right) \ .
\end{equation}
If we concentrate on the gravitational anomaly, encoded in $\CT$,
the inflow above completely cancels the one-loop contribution.
When we consider $n$ M5-branes, this inflow grows linearly with $n$,
and so is capable of canceling the gravitational anomaly from $n$ tensor
multiplets, also.

However, as is clear from the above, the cancelation
is not actually complete when we consider the axial anomaly
as well. Inflow $-I_8$ plus the one-loop anomaly $\CJ_8$
leave
\begin{eqnarray} \label{leftover}
2\pi(\CJ_8-I_8({\cal T}\oplus{\cal N})) =  2\pi \times\frac{1}{24}\,p_2({\cal N})
\end{eqnarray}
uncanceled \cite{Witten}. We will come back to how this remaining axial
anomaly is canceled shortly, as this mechanism is more subtle
and its variant will be needed  to clarify the D-brane anomaly inflow
in the next section. For now, let us first consider a slightly different
anomaly inflow to D-branes, where the bilinear and quadratic
inflows can be generated.

\subsection{I-Brane/D-Brane Inflow}

A more involved example of the anomaly inflow arises in the D-brane
context. The axial and gravitational anomaly are quite prevalent
and in fact most supersymmetric Yang-Mills theories with $d\ge 4$ have
such anomalies. Many of these theories are realizable as world-volume theories
from D-branes and Orientifold planes, whereby one must ask what
are the analog of the above anomaly inflow mechanism for D-branes.
On D$p$-branes, there are well-known topological
couplings between Ramond-Ramond tensor fields and the spacetime
curvature.
In fact, these coupling (modulo the normal bundle part) was conjectured
initially \cite{GHM} by asking that the anomaly associated with bi-fundamental
hypermultiplets along the intersections of two types of D-branes, with world-volumes
$M_1$ and $M_2$ respectively. Each set of D-branes carry the above
couplings, which can induce inflow onto the intersection $N=M_1\cap M_2$
and cancel anomaly due to the bi-fundamental fields there.
This is what is known as the I-brane anomaly inflow, with ``I" signifying
the intersection of D-branes. We will shortly review how this works
in a more general setting, including the case of $N=M_1=M_2$ \cite{CY},
for which case we refer to the D-brane inflow.

To produce the right anomaly inflow, one usually starts with
the world-volume topological coupling \cite{CY},
\begin{equation}
\sum_A (S_{CS}')^A\ ,
\end{equation}
where we summed over stacks of coincident D-branes, labeled by $A$, with the
revised topological coupling alluded to in (\ref{CS0'})\footnote{With our choice of unit $2\pi\sqrt{\alpha'}=1$, for D-brane discussions,
 $\mu_p={2\pi}/(4\pi^2\alpha')^{(p+1)/2}=2\pi$ and $2\kappa_{10}^2=(4\pi^2\alpha')^4/2\pi=1/2\pi$.
The curvature tensors in the topological couplings then have the standard normalization,
\begin{equation}
ch({\cal F})=\tr\, e^{{\cal F}/{2\pi}}\
\end{equation}
and so on.}
\begin{equation}
\label{CS3}
S_{CS}'=\frac{\mu_p}{2}  \int_{Dp}\;\left( s^*(C_{p+1})\wedge Y_0+(-1)^\epsilon \sum_{r<p} s^*(H_{r+2})\wedge
Y^{(0)}_{p-r-1}\right)\ ,
\end{equation}
with  $\epsilon=0,1$ for type IIA/IIB branes and
$H=dC+\cdots$.  We will keep track of the different
stacks, by labeling various world-volume objects,
such as $s^*$ or $Y$'s by the label $A$.
So, for example the characteristic classes are defined
as
\begin{equation}
Y^A_{n}\equiv  \left.\left[ch(\CF_A)\wedge \sqrt{\frac{\CA(\CT_A)}{\CA({\CN_A})}}\;\right]\right\vert_{n}
=\delta_n^0 \cdot Y_0^A+(1-\delta_n^0)\cdot d(Y^A)^{(0)}_{n-1}\ ,
\end{equation}
while the corresponding Chern-Simons densities $(Y^A)^{(0)} $ are
defined by $d((Y^A)^{(0)})=Y^A$.
The world-volume gauge field strength $\CF_A$
is in the fundamental representation of $U(Y_0^A=N_p^A)$.
The orientation of the world-volume
will be declared later when we discuss the equation
of motion for $C$'s.

Note that this form of the Chern-Simons couplings differ
from the natural world-volume topological couplings (\ref{CS2}),
and, as we will see shortly, the two generate two different inflow
even though the shifted Bianchi identities are the same:
Seemingly, only $S_{CS}'$ generates
the right inflow to cancel the world-volume one-loop anomaly,
which we will later attributes to mishandling of the Bianchi identity.

Note that the unfamiliar but crucial  factor $1/2$ in front
of the coupling. This reflects the subtlety  \cite{Deser}\cite{CY}
that we must include $C_{s}$  and their magnetic dual
$C_{8-s}$  on equal footing. In order for this to make
sense, the accompanying kinetic action for the RR fields must be
written in a way that does not distinguish electric and magnetic
fields, which effectively absorbs half of the usual minimal couplings.
An important consistency check is that this factor
$1/2$ does not appear in the field equations and Bianchi identities.
See Appendix A for a toy example that illustrates how this
is achieved, and Appendix B for precise form of the RR field
kinetic terms.

The equation of motion that follows from this coupling is
\begin{equation}
d\left( *(H_{r+2})\right)=-(-1)^r  \sum_{B} 2\kappa_{10}^2\mu_q \;
Y^B_{q-r}\wedge \Delta_{9-q}^B\ ,
\end{equation}
with some ``delta function" $(9-q)$-form, $\Delta_{9-q}^B$,
representing the D-brane position.
Because this is not
a scalar object, however, the expression becomes ill-defined
unless we carefully regularize and covariantize it.  This smearing
of the magnetic source is a recurring and necessary step when
we discuss the anomaly inflow, especially when the anomaly associated
with normal bundle needs to be discussed.
Thus, we write instead,
\begin{equation}\label{I-Bianchi}
d\left( *(H_{r+2})\right)= -(-1)^r\sum_{B} 2\kappa_{10}^2\mu_q \;
Y^B_{q-r}\wedge \tau_{9-q}^B\ ,
\end{equation}
where we smeared the sources due to the D$q$-branes by introducing
a ``delta-function"  $(9-q)$-form $\tau_{9-q}^B$, well-identified
in the mathematical literatures as  the Thom class of the normal
bundle $\CN$ \cite{BottTu}. See Appendices A and B for detailed
derivations.

We will
study it in more detail later, but it suffices to note here the
general form,
\begin{equation}
\tau_{9-q}=d(\rho\, \hat e_{8-q})\ .
\end{equation}
The ``radial" function $\rho$, whose support determines the
smearing of the source, interpolates between $-1$
on the brane and $0$ at infinity. The global angular
form $\hat e_{8-q}$  is essentially a covariantized
volume-form, normalized to unit volume, of a $(8-q)$-sphere
surrounding the D$p$-brane. In particular $\delta \hat e_{8-q}=0$,
and $d\hat e_{8-q}=0$ for even $q$ and $d\hat e_{8-q}=-\chi(\CN)_{9-q}$
with the Euler class $\chi$ for odd $q$. By choosing $\rho$ to have
increasingly small support near the origin, we can localize the
source with arbitrary precision, and with diffeomorphism invariance
preserved. In addition we will also choose $\rho'(0)=0$. With arbitrary
small support of $\rho$, we can take $Y$'s to be uniform along the
normal direction, which allows (\ref{I-Bianchi}) to make sense.

Since this equation of motion exists for all $C_{q+1}$'s, it also
implies, with $*H_n=(-1)^{(n-2+\epsilon)/2}H_{10-n}$, the modified Bianchi
identities
\begin{equation}\label{Bianchi-I}
d H_{8-r}= -\sum_{B} 2\kappa_{10}^2\mu_q \;(-1)^{(-q+\epsilon)/2}\wedge
\bar Y^B_{q-r}\wedge \tau_{9-q}^B\ ,
\end{equation}
with $\bar Y$'s being the complex conjugated $Y$'s,
\begin{equation}
\bar Y^A_{n}=\left. \left[ch(-\CF_A)\wedge
\sqrt{\frac{\CA(\CT_A)}{\CA({\CN_A})}}\;\right]\right\vert_{n}\ .
\end{equation}
We note here again that this shifted Bianchi implies that $S_{CS}$ of
(\ref{CS2}) and $S_{CS}'$ of (\ref{CS3}) are not equivalent. From
this, CY noted the following solution to the Bianchi
\begin{equation}\label{H}
H_{8-r}=d(C_{7-r}) -\sum_{B} 2\kappa_{10}^2 \mu_q\;
(-1)^{(-q+\epsilon)/2} (\bar Y^B)^{(0)}_{q-r-1}\wedge \tau_{9-q}^B\ ,
\end{equation}
and that gauge-invariance of the field strength
is ensured if they allow $C$'s to be gauge-variant as
\begin{equation}\label{delta'}
\tilde\delta C_{7-r} = \sum_{B} 2\kappa_{10}^2\mu_q \; (-1)^{(-q+\epsilon)/2}
(\bar Y^B)_{q-r-2}^{(1)}\wedge\tau_{9-q}^B\ ,
\end{equation}
where $\tilde\delta$  denotes the gauge transformation here,
to distinguish it against the revised one in section 4.
Thus $S_{CS}'$ is gauge-variant and
generates tree-level anomaly,
\begin{equation}
\tilde\delta S_{CS}'=\frac12 \sum_A \mu_p\int_A
\;\left( s_A^*(\tilde\delta C_{p+1}) Y^A_0+(-1)^\epsilon \sum_{r<p} s_A^*(H_{r+2})\wedge
d(Y^A)^{(1)}_{p-r-2}\right)\ .
\end{equation}
This is CY's master formula to the I-brane inflow, which has
been used to cancel many of known one-loop anomalies for field theories
on the intersecting brane. When we consider a pair of intersecting D-brane
stacks, this can cancel the anomaly from the bi-fundamental fermions propagating
along the intersection, in particular, which is the origin of the name
I-brane inflow.

A special case of this discussion occurs for a single stack of
D$p$-branes with $p= 5,7$.\footnote{$p=4, 6, 8$ are also acceptable,
except that the relevant field theories are of odd dimensions and
neither one-loop anomaly nor anomaly inflow is generated. $p=3$ appears
difficult since the product of two $\tau$'s will give 12-forms
and thus vanishes against  the spacetime integration. However, as we will
see later, this comes from mishandling of the Thom class
in this context.  }
The world-volume theories would be the maximally supersymmetric Yang-Mills theories in
$d=6,8$, whose one-loop anomaly polynomial is given in (\ref{one-loop}).
The gauge variations itself involves a factor of $\tau_{9-p}$,
which must be pulled-back to the world-volume defined by a limit of
the same $\tau_{9-p}$. While naively this looks like an ill-defined
procedure, this is not so because all the troublesome pieces in $\tau_{9-p}$
actually vanishes upon the pull-back, $s^*$, and the only surviving piece is
\begin{equation}
s^*(\tau_{9-p})=s^*(d\rho\wedge  \hat e_{8-p}-\rho\cdot \chi(\CN)_{9-p})=\chi(\CN)_{9-p}\ ,
\end{equation}
where we used $\rho'(0)=0$ and $\rho(0)=-1$. That is, the pull-back of the Thom class
to the zero section equals the Euler class \cite{BottTu}.

With  $1/2\times 2\kappa_{10}^2\mu_p^2=\pi$,
the anomaly inflow  $\tilde\delta S_{CS}'$ from the
self-intersection of these D$p$-branes is then
\begin{eqnarray} \label{adjoint}
&&(-1)^{(-p+1)/2}\pi \\
&&\times\int_{{\rm D}p}
\left( (\bar Y)^{(1)}_{2p-8}Y_0 + \sum_{6-p<r < p} \bar Y_{p+r-6} \wedge (Y^{(1)})_{p-r-2}
+ \bar Y_0Y^{(1)}_{2p-8}\right) \ ,
\nonumber
\end{eqnarray}
which equals, up to local counter terms,
\begin{eqnarray}
(-1)^{(-p+1)/2}\pi \int_{{\rm D}p}
\left(\sum_{r\ge 0} Y_r \wedge \sum_{s\ge 0} \bar Y_s \right)^{(1)}_{2p-8}
\wedge \chi(\CN)_{9-p}\ .
\end{eqnarray}
Using the definition of $Y$ and $\bar Y$'s and also
$ch_{\rm adj}^{SU(n)}(\CF)+1=ch(\CF)ch(-\CF)$
for $U(n)$ gauge group,  we find
\begin{eqnarray}
\tilde\delta S_{CS}'&=& - (-1)^{(p+1)/2 }\pi \int_{{\rm D}p}
\left([ch_{\rm adj}^{SU(n)}(\CF)+1]\wedge \frac{\CA(\CT)}{\CA({\CN})} \right)^{(1)}
\wedge \chi(\CN)\ .
\end{eqnarray}
When $p\ge 5$, this is
equivalent to, again up to local counter-terms,\footnote{
This last step works because for $p\ge 5$ the 0-form part
of the characteristic classes, $ch$ and $\CA$, are irrelevant
upon integration; $\chi$ is a $(9-p)$-form and the integration
is over $(p+1) > (9-p)$ dimensions.}
\begin{eqnarray}
\tilde\delta S_{CS}'&=& - (-1)^{(p+1)/2 }\pi \int_{{\rm D}p}
\left([ch_{\rm adj}^{SU(n)}(\CF)+1]\wedge \frac{\CA(\CT)}{\CA({\CN})} \wedge \chi(\CN)\right)^{(1)}
\ ,
\end{eqnarray}
 and equals, upon the identity
$
{\chi(\CN)} {\CA({\CN})}^{-1}=ch_+(\CN)-ch_-(\CN)
$,
\begin{eqnarray}
= - (-1)^{(p+1)/2}\pi \int_{{\rm D}p}
\left([ch_{\rm adj}^{SU(n)}(\CF)+1]\wedge{\CA(\CT)}\wedge[ch_+(\CN)-ch_-(\CN)]  \right)^{(1)}\ .
\end{eqnarray}
which has precisely the right form to cancel the one-loop anomaly
(\ref{one-loop}) of the maximally supersymmetric $U(n)$ Yang-Mills
theory in the respective dimensions.

By the way, the overall sign is not related to whether we
are considering D$p$'s or anti-D$p$'s.
For a single stack, an extra overall sign in the coupling of
D-branes to $C$'s, cancels out when we put back $\tilde\delta C$
to compute variation of $S_{CS}'$. From one-loop perspective,
this happens because, as we flip the chirality of world-volume
fermions, their representations under $SO(9-p)$ R-symmetry also
flip as the fermions have a definite ten-dimensional chirality;
The sign flip from the chirality flip is canceled by the exchange
of $S_+(\CN)$ and $S_-(\CN)$ representations. As was mentioned in
footnote $\#$2, this overall sign appears to be associated with
the canonical choice of the chirality operator and the
accompanying signature $(-++\cdots+)$, relative to
those chosen in Ref.~\cite{AlvarezGaume:1983ig}.

There are a couple of unsatisfactory issues that
remain here. One problem, as mentioned several times already,
concerns the case of $p=3$, which apparently produces no
inflow. This is due to $S^*(\tau_6)=\chi_6$
in the inflow formula, since a 6-form integrates to zero against
the four world-volume dimensions. The one-loop axial anomaly is
nontrivial, and something else must compensate for the anomaly,
yet it is difficult to imagine D3-branes,
despite their self-dual nature, can be that different.

The second issue, which is a little more of technical nature,
is that, to produce the correct inflow for generic cases, one must
use $S_{CS}'$  instead of $S_{CS}$. Although the resulting Bianchi
identity is the same, the action themselves are not equivalent, and
one finds
\begin{equation}\label{diff}
\tilde\delta S_{CS}'\neq \tilde\delta S_{CS}\ ,
\end{equation}
even up to local counter terms. If we started
with $S_{CS}$ and followed the same procedure as above,
we would have arrived at an analog of (\ref{adjoint})
effectively without the last term in the parenthesis there.
$S_{CS}$ in (\ref{CS2}) looks far more natural, but does not
yield the right canceling inflow even for $p=5,7$. Both of these
curiosities were noted by Cheung and Yin \cite{CY}.

As we will see in section 4, these two problems have a common origin and is solved by
more careful treatments of the regularized source $\tau$'s.

\section{M5-Brane Axial Anomaly Inflow}

A slightly simpler version of this last issue has been discussed
in the context of M5-brane normal bundle anomaly, so we will
review this first. Recall that, after the anomaly inflow from $G_4\wedge I_7^{(0)}$
term onto a M5-brane, we have a leftover
\begin{equation}
2\pi\times\frac{1 }{24}\,p_2(\CN)\ ,
\end{equation}
as shown in (\ref{leftover}). Further cancelation of this is more subtle
and known to originate from a  revised version of the
spacetime Chern-Simons coupling
\begin{equation}
-\frac{2\pi }{6}\int C_3\wedge G_4\wedge G_4\ ,
\end{equation}
upon careful regularization of the $C_3$ \cite{FHMM}. Here we follow
FHMM \cite{FHMM} almost verbatim.\footnote{Except for renaming the
global angular form as $$\hat e_4=(e_4/2)_{FHMM}$$ and
clarification of a  related normalization issue.}

To obtain the right inflow, let us recall the Bianchi identity (\ref{Bianchi})
(with  $2\pi l_p=1$)
\begin{equation}
dG_4 =
\delta(y^1)\cdots\delta(y^5)dy^1\cdots dy^5\ ,
\end{equation}
with normal bundle coordinate $y^i$'s. As in the I-brane/D-brane discussion,
this expression needs modification if we wish to be careful about the
normal bundle part. We should substitute the right hand side with a covariant
and smeared version of the delta function, namely
the Thom class
\begin{equation}
dG_4 =\tau_{M5}({\cal N})\ ,
\end{equation}
which can be written as before
\begin{equation}\label{dG_4}
\tau_{M5} = d
\big[\rho(r)\wedge \hat e_4\big ]= d\rho\wedge \hat e_4\ .
\end{equation}
$\rho (r)$ is a smooth function of radial direction, with
$d\rho$ serving as a smoothed radial delta-function
satisfying $\rho(r)=-1$ on the M5-brane and $\rho(r)=0$
far from the branes. $\hat e_4$ is a global angular form, which
is closed as the normal bundle is of odd dimension.\footnote{
An interesting attempt to assign  a microscopical origin of
such a smearing, albeit in a toy model, can be found in Ref.~\cite{Boyarsky:2002ck}.}

More
explicitly, we have
\begin{eqnarray}\nonumber
\hat e_4(\Theta) &=& \frac{1}{64\pi^2}\epsilon_{a_1\cdots a_5}
\big[(D\hat y)^{a_1}(D\hat y)^{a_2} (D\hat y)^{a_3}(D\hat y)^{a_4}\hat y^{a_5}\\
&&-2F^{a_1a_2}(D\hat y)^{a_3}(D\hat y)^{a_4} y^{a_5} + F^{a_1a_2}F^{a_3a_4}\hat y^{a_5}\big]\ ,
\end{eqnarray}
with
\begin{eqnarray}
(Dy)^a = d\hat y^{a}-\Theta^{ab}\hat y^b\ ,\quad
F^{ab} = d\Theta^{ab}-\Theta^{ac}\wedge\Theta^{cb}\ ,
\end{eqnarray}
in terms of $SO(5)$ connection $\Theta^{ab}=-\Theta^{ba}$
and the normalized Cartesian coordinates $\hat y^a=y^a/r$
along the fibre. Using
$\delta \Theta^{a_1a_2} = (D\Lambda)^{a_1a_2}$ and
$\delta \hat y^a = \Lambda^{a_1a_2}\hat y^{a_2}$, the descents of
$e_4$,
\begin{equation}
\hat e_4 =d\hat e_3^{(0)},~~\delta \hat e_3^{(0)}=d\hat e_2^{(1)},
\end{equation}
are
\begin{eqnarray}\nonumber
\hat e_3^{(0)}(\Theta) &=& \frac{1}{32\pi^2}\epsilon_{a_1\cdots a_5}
\big[ \Theta^{a_1a_2}d\Theta^{a_3a_4} \hat y^{a_5}\\
&& -\frac12 \Theta^{a_1a_2}\Theta^{a_3a_4}d\hat y^{a_5}
-2\Theta^{a_1a_2}d\hat y^{a_3}d\hat y^{a_4}\hat y^{a_5}\big]\ ,
\end{eqnarray}
and
\begin{equation}
\hat e_2^{(1)} (\Lambda, \Theta) = \frac{1}{16\pi^2}\epsilon_{a_1\cdots a_5}
\big[\Lambda^{a_1a_2}d\hat y^{a_3}d\hat y^{a_4}\hat y^{a_5}
-\Lambda^{a_1a_2}\Theta^{a_3a_4}d\hat y^{a_5}\big]\ .
\end{equation}

Now we can solve the Bianchi identity (\ref{dG_4}),
\begin{equation}
G_4 = dC_3 +
\big[\beta\rho \hat e_4 -(1-\beta) d\rho\wedge \hat e_3^{(0)}\big]\ ,
\end{equation}
with arbitrary real number $\beta$. Note that
$\rho e_4$ diverges at the origin, since integral over any arbitrary small
four-sphere around the origin gives a finite value. On the other hand,
$d\rho\wedge \hat e_3^{(0)}$ can be managed to be finite near the origin, by requiring
$d\rho \rightarrow 0$ as $r\rightarrow 0$. Hence we should choose $\beta=0$, to
ensure the regularity of $C_3$ and $G_4$ near the M5-branes, so
\begin{equation}\label{T-descent}
G_4 = dC_3 -  d\rho\wedge \hat e_3^{(0)}\ .
\end{equation}
The solution implies that $C_3$ transforms nontrivially under the $SO(5)_R$
gauge transformation in order that $G_4$ is invariant:
\begin{equation}
\delta C_3 = -d\rho \wedge \hat e_2^{(1)}\ .
\end{equation}

In the presence of such M5 sources, the Chern-Simons term
\begin{equation}
-\frac{2\pi }{6} \int_{M_{11}} C_3\wedge G_4\wedge G_4\ ,
\end{equation}
becomes ambiguous. FHMM suggested that the right modification
is to replace $C_3$ by   $C_3-\sigma_3\equiv C_3-\rho\hat e^{(0)}_3$
and $G_4$ by $G_4 -\rho \hat e_4$ with the properties,
\begin{eqnarray}\nonumber
G_4 -\rho \hat e_4 &=& d(C_3 -\sigma_3)\ ,\\
\delta(C_3 -\sigma_3) &=& d(-  \rho \cdot \hat e_2^{(1)})\ .
\end{eqnarray}
The modified Chern-Simons term
\begin{equation}
S'_{CS} = -\frac{2\pi}{6}\lim_{\epsilon\rightarrow 0}\int_{M_{11}-D_\epsilon(M5)}
(C_3 -\sigma_3)\wedge d(C_3 -\sigma_3)\wedge d(C_3 -\sigma_3)\ ,
\end{equation}
where we subtract the infinitesimal tubular neighborhood,  $D_{\epsilon}(M5)$,
of the world-volume $M5$ with arbitrary small radius $\epsilon$.
Its gauge-variation is
\begin{equation}
\delta S'_{CS} = \frac{2\pi}{6}\lim_{\epsilon\rightarrow 0}\int_{M_{11}-D_\epsilon(M5)}
d(\rho\cdot \hat  e_2^{(1)})\wedge d(C_3-\sigma_3)\wedge d(C_3-\sigma_3)\ .
\end{equation}
of which $C_3$ parts vanish with $\epsilon\rightarrow 0$. The integrand is
a well-defined total derivative, so we are left with an integral over the
sphere bundle $S_\epsilon(M5)$ of vanishing radius
\begin{equation}
\delta S'_{CS} = -\frac{2\pi 
}{6}\int_{S_\epsilon(M5)}{\hat e_4}\wedge {\hat e_4}\wedge{\hat e_2^{(1)}} = -2\pi 
\int_{M5} \frac{p_2({\cal N})^{(1)}}{24}\ ,
\end{equation}
which neatly cancels the normal bundle anomaly that was left over in
section 2.1.

The two anomaly inflows to the M5-brane can each be generalized  easily
to the case of $N$ coincident branes. One inflow is linear in $C$ while
the other is cubic in $C$, so  the total inflow has to be
\begin{equation}
-2\pi \left(N\times I_8(\CT\oplus \CN)+N^3\times\frac{1}{24}\,p_2(\CN)\right)\ ,
\end{equation}
from which we infer the world-volume one-loop anomaly of $A_{n-1}$ $(2,0)$
theory plus a free tensor multiplet theory as
\begin{equation}
2\pi \left((N-1)\times \CJ_8(\CT, \CN)+(N^3-N)\times\frac{1}{24}\,p_2(\CN)\right)+2\pi
\CJ_8(\CT, \CN)\ .
\end{equation}
Of course this shows the famous $N^3$ scaling of $(2,0)$ theories \cite{hep-th/9806087}\cite{Harvey:1998bx}\cite{Intriligator:2000eq}\cite{Yi:2001bz}.

\section{D-Brane Anomaly Inflow Revisited}

The salient point we wish to learn from the M5-brane axial anomaly inflow
is how, in (\ref{T-descent}),  FHMM solved the Bianchi identity
in the presence of  a smeared delta
function source in the form of the Thom class. Even though the Thom
class on the M5-brane was introduced as $d(\rho\cdot \hat e_4)$, the
descent formula $\tau_{M5}=d\tau_{M5}^{(0)}$ was written as
\begin{equation}
\tau^{(0)}_{M5}=-d\rho\wedge \hat e_3^{(0)}\ ,
\end{equation}
$\tau^{(1)}_{M5}$ of which at the end generated the necessary inflow
for the M5-brane axial anomaly.
The argument in favor of $-d\rho\wedge \hat e_3^{(0)}$ as the unique
choice (instead of $\rho\cdot \hat e_4$) is that, since one resolved
the magnetic source into a smooth configuration, the field strength
should remain smooth everywhere and in particular on the
M5-brane. With $\rho(0)=-1$, $\rho \hat e_4$ is
singular and ill-defined at the origin, whereas $-d\rho\wedge \hat e_3^{(0)}$
is regular at the origin as long as we choose $\rho'(0)=0$.

Note that, in the discussion of I-brane/D-brane inflow, the
shifted Bianchi identity (\ref{I-Bianchi}) also
involved Thom classes $\tau_{9-q}=d(\rho\cdot\hat e_{8-q})$
for the D$q$-branes, but was solved  as
\begin{equation}
H_{s+2}= \cdots -\sum_{B} 2\kappa_{10}^2\mu_q \;(-1)^{(-q+\epsilon)/2}
(\bar Y^B)^{(0)}_{q+s-7}\wedge \tau_{9-q}^B\ .
\end{equation}
Note that the descent of $\tau_{9-q}$ is apparently not invoked.
Actually, for terms with $q+s=6$, for which $\bar Y$ factor is a number,
$\tau^{(0)}_{8-q}$ must appear on the right hand side,
since otherwise the Bianchi identity is not obeyed.
Presumably this term is  suppressed because the obvious choice
$\tau^{(0)}=\rho \cdot\hat e$ is gauge-invariant and seemingly irrelevant
for the inflow. However, $\rho\cdot \hat e$
is neither a unique choice for descent nor physically sensible.
With $\rho=-1$,
$H_{s+2}\sim \cdots+ \bar Y_0\cdot\rho\cdot \hat e_{s+2}+\cdots$
would be singular and ill-defined at the origin.
In this section, we will address this problem and study the ramifications.

We wish to emphasize here that we will be using, instead of $S_{CS}'$
of (\ref{CS3}), the natural Chern-Simons coupling (\ref{CS2}) which is
\begin{equation} \label{CS4}
 S_{CS}=\frac{\mu_p}{2} \,\int_{Dp} \;\sum_{r\le p} s^*(C_{r+1})\wedge Y_{p-r}
\ ,
\end{equation}
with $Y= ch(\CF)  \CA(\CT)^{1/2}\CA({\CN})^{-1/2}$.
As usual, $s^*$ is the pull-back to the world-volume (i.e., to the zero
section of the normal bundle). With the revised inflow mechanism below,
$S_{CS}$ and $S_{CS}'$ will be shown to generate the equivalent anomaly inflow
in the end.

\subsection{Revised Inflow from Regularity of $H$ }\label{Inflow}

First, we need to clarify an important difference between the
Thom classes of even and odd dimensional bundles. For
odd fibre dimensions (applicable to even $q$ and thus type IIA branes),
$\tau_{9-q}=d(\rho\cdot\hat e_{8-q})$
behaves in much the same way as $\tau_{M5}$ of the previous section. For
even fibre dimensions (applicable to odd $q$ and thus type IIB branes),
the global angular form decomposes into two pieces \cite{BottTu}\cite{BB}
\begin{equation}
\hat e_{8-q}=v_{8-q}+\Omega_{8-q}(\CN)\ ,
\end{equation}
where the first term involves at least one normal vector field $\hat y$
and can be written locally as
\begin{equation}
v_{8-q}=d\psi_{7-q}\ ,
\end{equation}
while the last term is nothing but the Chern-Simons term of
the Euler class with a sign flip, i.e.,
\begin{equation}
d\Omega_{8-q}(\CN)=-\chi(\CN)_{9-q}\ .
\end{equation}
Clearly, this behavior of the Thom class is responsible,
with $\rho(0)=-1$, for the identity $s^*(\tau)= \chi$.
Finally the gauge-invariance of $\hat e$ implies that
\begin{equation}
\delta\psi_{7-q}=-\Omega^{(1)}_{7-q}=\chi(\CN)^{(1)}_{7-q}\ .
\end{equation}
$\Omega$ exists for even-dimensional normal bundles, and
so this is relevant for all type IIB branes.

Note that $v_{8-q}$ (and its descent $\psi_{7-q}$) is
singular at the origin, being a normalized volume form of
$S^{8-q}$. In contrast, $\Omega(\CN)_{8-q}$ is composed only of
the gauge fields of the normal bundle and is well-defined and smooth
everywhere. For regular solutions of $H$, we must then choose the
following descent for $\tau$,
\begin{equation} \label{tauzero}
\tau^{(0)}_{8-q}=-d\rho\wedge\psi_{7-q} +\rho\cdot \Omega_{8-q}\ ,
\end{equation}
which results in
\begin{equation}
\tau^{(1)}_{7-q}=-\rho\cdot \chi(\CN)^{(1)}_{7-q}\ .
\end{equation}
Note that both expressions are regular at the origin, with $\rho'(0)=0$.
This gives
\begin{equation}
H_{s+2}=d(C_{s+1})- \sum_{B} 2\kappa_{10}^2 \mu_q\;
(-1)^{(-q+\epsilon)/2} \left(\bar Y^B\wedge\tau^B\right)^{(0)}_{s+2} \ ,
\end{equation}
where, for type IIB theory,
\begin{equation}
\left(\bar Y^B\wedge\tau^B\right)^{(0)}_{s+2}
=
\beta(\bar Y^B)_{q+s-7}^{(0)}\wedge\tau_{9-q}^B
+(1-\beta) (\bar Y^B)_{q+s-6}\wedge(-d\rho\wedge\psi_{7-q}+\rho\cdot\Omega_{8-q})^B \ .
\end{equation}
Although $\beta$ is an arbitrary real number in general, we must take
$\beta=0$ when $\bar Y$ on the left hand side is a 0-form (here, $q+s=6$).
Its gauge variation gives
\begin{eqnarray}
\left(\bar Y^B\wedge\tau^B\right)^{(1)}_{s+1} =\beta (\bar Y^B)_{q+s-8}^{(1)}
\wedge \tau^B_{9-q}+ (1-\beta) \bar Y^B_{q+s-6}\wedge(-\rho\cdot\chi^{(1)}_{7-q})^B\ .
\end{eqnarray}
With this understood, the gauge transformation of $C$ is,
\begin{equation}\label{new}
\delta C_{s+1}=   \sum_{B} 2\kappa_{10}^2 \mu_q\;
(-1)^{(-q+\epsilon)/2} \left(\bar Y^B\wedge \tau^B\right)^{(1)}_{s+1}\ .
\end{equation}
Note the difference from (\ref{delta'}). The difference is
essential for terms with $\bar Y_{0=q+s-6}$ factor and its
consequence in $\delta S_{CS}$ below cannot be removed by local
counter-terms.

Let us concentrate on the case of a single stack of type IIB
D$p$-branes. The gauge variation of $S_{CS}$  (\ref{CS4}) is
\begin{equation}
\delta S_{CS}=  (-1)^{(-p+1)/2 }\pi  \int_{Dp}   \sum_{r} s^*\left(\left(\bar Y_{p+r-6}
\wedge\tau_{9-p}\right)^{(1)}\right)\wedge Y_{p-r} \ .
\end{equation}
Just as $s^*(\tau)=\chi$,
it is easy to show that
\begin{equation}
s^*(\tau^{(1)})=s^*(-\rho\chi^{(1)})=\chi^{(1)}\ ,
\end{equation}
and that
\begin{equation}\label{variation}
\delta S_{CS}=  (-1)^{(-p+1)/2 }\pi  \int_{Dp}   \sum_{r} \left(\bar Y_{p+r-6}
\wedge\chi_{9-p}(\CN)\right)^{(1)}\wedge Y_{p-r}\ ,
\end{equation}
which equals
\begin{equation}
- (-1)^{(p+1)/2 } \pi \int_{{\rm D}p}  \left( ch(-\CF)  \wedge\sqrt{\frac{\CA(\CT)}{\CA({\CN})}} \wedge
\chi(\CN)\right)^{(1)}\wedge
\left( ch(\CF)\wedge   \sqrt{\frac{\CA(\CT)}{\CA({\CN})}} \right)\ .
\end{equation}
With $p<9$,  $\chi(\CN)_{9-p}$ is never 0-form,  allowing us to rewrite this as,
up to local counter terms,\footnote{$p=9$ requires a separate discussion since this case
involves Orientifold planes. See next section.}
\begin{eqnarray}
\delta S_{CS}
&=&- (-1)^{(p+1)/2 } \pi \int_{{\rm D}p}  \left(  ch(\CF)\wedge ch(-\CF)
\wedge \frac{\CA(\CT)}{\CA({\CN})} \wedge \chi(\CN)\right)^{(1)} \\ \nonumber \\
&=& - (-1)^{(p+1)/2}\pi \int_{{\rm D}p}
\left([ch_{\rm adj}^{SU(n)}(\CF)+1]\wedge{\CA(\CT)}\wedge[ch_+(\CN)-ch_-(\CN)]  \right)^{(1)}\ .\nonumber
\end{eqnarray}
Of these, for $p=1$,
the expression is null and no inflow is generated. For others, $p=3,5,7$,
this is precisely the right inflow to cancel one-loop anomaly (\ref{one-loop})
for $d=4,6,8$. ($p=9$ requires a special treatment as it always involves
an Orientifold plane. See section 5.)

We have re-analyzed the Bianchi identities of RR field strengths
by requiring the regularity of physical variables. This is not by a choice but
required, since the D-brane inflow analysis must have the magnetic sources
regulated anyway. Singular field strengths in the absence of singular
source do not make any sense. Although the final answer looks superficially
the same as before, it differs in two important aspects and addresses the
concerns raised at the end of section 2.2.
First, the revised inflow is now applicable for all D-branes and I-branes.
In particular, it produces right answers for a single stack of D$p$-branes
including $p= 3$ case while the old procedure produced a null result for
$p=3$ and produced right answers for $p\ge 5$ only.
Second, with the revised gauge transformation rule, $\delta\neq\tilde \delta$,
the natural Chern-Simons couplings  $S_{CS}$ (\ref{CS2}) generates
the correct anomaly inflow. Furthermore, it is not difficult to see that,
{\it up to local counter terms on world-volume},
\begin{equation}
\delta S_{CS}=\delta S_{CS}' \ ,
\end{equation}
so the  unreasonable sensitivity to the precise form of Chern-Simons
couplings, as in (\ref{diff}), is no longer there.

\subsection{Axial Anomaly Inflow onto D3-Branes}

In particular, this  resolves a long-standing issue regarding
D3-brane anomaly inflow. Our analysis shows that the correct
inflow arises also for D3-branes; one-loop anomaly of the maximal $U(N_3)$
super Yang-Mills theory is completely canceled by the anomaly inflow onto $N_3$
coincident D3-branes.
Previous analysis produced a null inflow for this case, seemingly
requiring another inflow mechanism \cite{CY}. The crucial difference between
the old and the revised inflow is whether one has a 6-form $s^*(\tau_6)=\chi_6$
as a blind overall factor (which kills off all terms) or one also has
an exceptional term with 4-form $s^*(\tau_4^{(1)})=\chi^{(1)}_4$ instead.
Here we wish to retrace the case of D3-branes, with
more care given to details of the Thom class, for a pedagogical reason.

Upon close
inspection of the inflow, one can see easily that, for D$p$-branes,
only those RR gauge fields from $C_{p+1}$ down to its dual $C_{7-p}$
contribute to the inflow. For an $N_3$ coincident D3, $C_4$ is self-dual,
 and the only relevant term for D3-brane inflow is the
  minimal coupling\footnote{
Since D3 is a self-dual object, one might wonder whether
this electric-type minimal coupling suffices. See Appendix B for why this is
so when one uses the duality symmetric formulation for RR-gauge field
kinetic term. The latter in particular imposes the self-duality condition
for $C_4$ as an equation of motion, with or without the minimal coupling to
$C_4$. }
\begin{eqnarray}
S_{CS}^{D3} = \frac{\mu_3N_{3}}{2}\int_{D3} s^*(C_4)  \ ,
\end{eqnarray}
with the constant 0-form $Y_0=N_3=\bar Y_0$.
This is also related to the fact that
$s^*( \tau^{(1)})=\chi^{(1)}$ is already a 4-form, saturating all the world-volume
dimensions.
From this, combining with the self-duality constraint on $C_4$,
we have the Bianchi identity of $H_5$
\begin{equation}
dH_5 = 2\kappa_{10}^2\mu_3 N_3 \tau_{6}(D3)\ ,
\end{equation}
again with the regularized and covariantized $\tau_6 (D3)$.

Recall that this
Thom class is defined by
\begin{equation}
\tau_6(D3) = d(\rho \cdot \hat e_5)\ ,
\end{equation}
with the global angular five-form $\hat e_5$ of unit volume. More explicitly,
\begin{eqnarray}\nonumber
\hat e_5 &=& -\frac{1}{15} \epsilon_{a_1\cdots a_6}D\hat y^{a_1}D\hat y^{a_2}D\hat y^{a_3}
D\hat y^{a_4}D\hat y^{a_5}\hat y^{a_6}\\
&& -\frac{1}{6}\epsilon_{a_1\cdots a_6} F_R^{a_1a_2}D\hat y^{a_3}D\hat y^{a_4}D\hat y^{a_5}
\hat y^{a_6} -\frac{1}{8}\epsilon_{a_1\cdots a_6} F_R^{a_1a_2}F_R^{a_3a_4}D\hat y^{a_5}
\hat y^{a_6}\ ,
\end{eqnarray}
which can be decomposed as
\begin{equation}
\hat e_5 = d\psi_4 + \Omega_5\ ,
\end{equation}
with
\begin{equation}
d\psi_4=-\frac{1}{120\pi^3}\; \epsilon_{a_1\cdots a_6}d\hat y^{a_1}d\hat y^{a_2}
\cdots d\hat y^{a_5}\hat y^{a_6} +\cdots\ ,
\end{equation}
and
\begin{equation}
\Omega_5 = \frac{1}{384\pi^3}\,\epsilon_{a_1 \cdots a_6}
\left[F_R^{a_1a_2}F_R^{a_3a_4} A_R^{a_5 a_6}+\cdots\right]\ ,\qquad
d\Omega_5
=-\chi_6(F_R)\ .
\end{equation}
Of course the six-form $\chi_6$ and the five-form $\Omega_5$ vanish identically
when evaluated on  the four dimensional world-volume of D3, but what matters at
the end is the appearance of the 4-form $\chi^{(1)}_4$ from the variation of $\psi_4$.
In what follows, we obtain the same final answer if we remove $\chi_6$ and $\Omega_5$
from all the formulae but remember that $\delta\psi_4$ is trivially closed on the
D3 world-volume.

As before, from the regularity requirement of $H_5$ and $C_4$,
we must choose among many naive choices of $[\tau_6(D3)]^{(0)}$,
\begin{equation}
H_5  = dC_4 +2\kappa_{10}^2\mu_3N_3\left(\tau(D3)\right)^{(0)}_5
=dC_4 +2\kappa_{10}^2\mu_3 N_3\left[\rho\wedge \hat{e}_5
- d(\rho\wedge\psi_4)\right]\ .
\end{equation}
On the other hand, 
since
\begin{equation}
\delta\hat e_5=0\ ,\qquad
\delta \psi_4 
= \chi_4^{(1)}\ ,
\end{equation}
the gauge invariance of $H_5$ yields
\begin{equation}
s^*(\delta C_4) = -2\kappa_{10}^2\mu_3 N_3\times s^*\left(\tau_6(D3)^{(1)}\right)
= -2\kappa_{10}^2\mu_3 N_3\times \chi_4^{(1)}\ .
\end{equation}
If we substitute this to $\delta S_{CS}^{D3}$, we finally have
\begin{eqnarray}\label{D3inflow}
\delta S_{CS}^{D3} = -\kappa_{10}^2\mu_3^2 {N_3^2}\int_{D3}\chi^{(1)}_4
=N_3^2\times \left(-\pi \int_{D3}\chi^{(1)}_4\right)\ ,
\end{eqnarray}
with $\kappa_{10}^2\mu_3^2 =[(2\pi)^7(\alpha')^4/2]\times [1/(2\pi)^3(\alpha')^2]^2=\pi $.
This cancels  exactly  the one-loop anomaly on the D3-branes.

As we saw in the introduction,
the SO$(6)_R$ axial anomaly polynomial at one-loop of
the $U(N_3)$ theory is
\begin{eqnarray}
I_6&=&\frac{N_3^2}{24\pi^2} \, {\rm tr}_{S^+} \,F_R^3
\;=\;N_3^2\cdot 2\pi\cdot ch_{S^+}(F_R)\biggl\vert_{6-form} \nonumber \\
&=&N_3^2\cdot \pi\cdot\left[ch_{S^+}(F_R)-ch_{S^-}(F_R)\right]\biggl\vert_{6-form}\ ,
\end{eqnarray}
where $F_R$ is the curvature tensor of an external $SO(6)_R$
in the Weyl representation. The bracket in the last line equals the Euler
class divided by the A-roof genus, and the Euler class is already
6-form, so the one-loop anomaly polynomial equals
\begin{equation}
I_6= N_3^2\times \pi \chi(F_R)\ ,
\end{equation}
which is  precisely canceled by the inflow (\ref{D3inflow}).

The case of D3 is special in that the minimal coupling to $C_4$ alone generates
the anomaly inflow and there is no need to invoke
lower-rank RR gauge fields. This happens due to the self-dual nature of D3.
A toy model of such self-dual objects, namely dyonic string in six dimensions, was studied
previously in Refs.~\cite{Henningson:2004dh}\cite{Berman:2004ew}\cite{Henningson}.
Our inflow argument is related most directly to that of Ref.~\cite{Berman:2004ew}.
There is also some relation to Ref.~\cite{Henningson:2004dh} in that $v_5=d\psi_4$
is the generalization of the Wess-Zumino-Witten term of the latter, but
the inflow here is a direct consequence of the standard topological
coupling, rather than with additional modifications. In particular,
the smearing function $\rho$ plays a crucial role here.

\section{Chern-Simons Couplings on Orientifold Planes}

Extending all of these to the presence of Orientifold planes
should be straightforward. The main extra ingredient is how
the various Orientifold planes couple to the space-time curvature.
For O$p^-$ plane, the relevant Chern-Simons coupling is known to be,
\begin{equation}\label{Ominus}
S_{Op^-} = \frac12\times \left(-2^{p-4}\frac{\mu_p}{2}\int_{Op^-}  \sum_r  s^*(C_{r+1})\wedge
\sqrt{\frac{{\cal L}({\cal T}/4)}{{\cal L}({\cal N}/4)}} \;\right)\ ,
\end{equation}
where $\cal L$ is the Hirzebruch class
\cite{hep-th/9812071}\cite{hep-th/9812088}\cite{hep-th/9707224}\cite{hep-th/9709219}.
There are various
studies in the past that worked out analog of this for other
three classes of Orientifold planes, but the answers seem
to disagree partially with one another \cite{DFM}\cite{Scrucca:1999jq}\cite{Mukhi}\cite{Henry}\cite{JFO}.

In this section, we will show that the one-loop anomaly from
the gauge sector cancels away by the anomaly inflow, if we assume
the most obvious choices of the Orientifold Chern-Simons couplings, which
in addition to the above O$^-$,
\begin{equation}\label{Otilde}
S_{\widetilde{Op^-}} = \frac12 \times\left( -\frac{\mu_p}{2}\int_{\widetilde{Op^-}}  \sum_r  s^*(C_{r+1})\wedge\left[
\sqrt{\frac{{\cal A}({\cal T})}{{\cal A}({\cal N})}}
-2^{p-4}
\sqrt{\frac{{\cal L}({\cal T}/4)}{{\cal L}({\cal N}/4)}}\right] \; \right)\ ,
\end{equation}
reflecting the usual statement that this case has a single,
unpaired D-brane stuck at the Orientifold plane. For O$p^+$,
\begin{equation}\label{Oplus}
S_{Op^+} =\frac12\times\left( 2^{p-4}\frac{\mu_p}{2}\int_{Op^+}  \sum_r s^*(C_{r+1})\wedge
\sqrt{\frac{{\cal L}({\cal T}/4)}{{\cal L}({\cal N}/4)}} \;\right)\ ,
\end{equation}
and the same expression for $S_{\widetilde{Op^+}}$.
This last one associated with symplectic type orbifolding
agrees with Refs.~\cite{DFM}\cite{Scrucca:1999jq}.

As before, the overall factor $1/2$  exists only when we write the kinetic terms of
RR tensors in the duality symmetric form, and does not enter the equation of
motion. The other 1/2 factor accompanying $\mu_p$ is due to the Orientifolding
projection.

\subsection{D$p$-O$p$ Inflow}\label{OInflow}

We work in the covering space of the Orientifold and take care
to divide by two at the end of everything. For example, equation
of motion and the Bianchi identity are unaffected by this, but the action written
in the covering space must be either divided by two (e.g., world-volume part)
or restricted to the half space (e.g., spacetime part).
Similarly, the D-brane Chern-Simons couplings are
\begin{equation}\label{S2}
S_{Dp} = \frac12\times \left(\frac{1}{2} \sum_A \mu_p \int_A \sum_{r\le p}s^*( C_{r+1})\wedge ch_{2k}({\cal F_A})
\wedge\sqrt{\frac{{\cal A}({\cal T}_A)}{{\cal A}({\cal N}_A)}} \;\right)\ .
\end{equation}
Note that here we assumed these $2k$ D$p$ branes are on the top
of the O$p^-$ plane, so they share the Thom class $\tau$, the tangent bundle $\cal T$,
and the normal bundle $\cal N$.

Note that, upon the Orientifold
projection, some of the RR tensor fields are absent. With
O$p$ planes, $C_{p-1\pm 4n}$ maps to
its negative and thus are projected out, while $C_{p+1\pm 4n}$
remains intact. This can potentially modify inflow argument.
However, we do not really lose any term since $ch_{2k}({\cal F})$
is a sum of $4n$-forms for $SO(2k)$ and $Sp(k)$ gauge groups,
and  since the Euler character $\chi_{9-p}$ is a $(9-p)$-form
monomial.
An exception to this is $p=9$, for which one of the
relevant RR gauge field, $C_{10=9+1}$, does not  exist,
and  $Y_{10}^{(1)}$ type of inflow cannot be generated.
This is precisely what leads to the
tadpole condition $2k=32$ for type I string theory.
See section (\ref{typeI}) for a separate review of D9-O9.

With this, we may proceed as before except that
$Y=ch_{2k}(\CF)\CA(\CT)^{1/2}\CA(\CN)^{-1/2}$ is shifted
by $-2^{p-4}$ times
\begin{equation}\label{definition}
Z\equiv \sqrt{\frac{{\cal L}({\cal T}/4)}{{\cal L}({\cal N}/4)}}\ ,
\end{equation}
and the Bianchi identity reads
\begin{equation}
d(H_{s+2}) =-\sum_B 2\kappa_{10}^2 \mu_q (-1)^{(-q+\epsilon)/2}
(\bar Y^{B}_{q+s-6}-2^{p-4}\bar Z_{q-r}^B)\wedge \tau_{9-p}^B\ ,
\end{equation}
from which we repeat the procedure of sec.\ref{Inflow} and
arrive at the world-volume expressions,
\begin{eqnarray}\nonumber
\delta(S_{Op^-}+S_{Dp}) 
&=& -(-1)^{(p+1)/2}\cdot\frac{\pi}{2}\int \left(\bar Y\wedge Y
\wedge \chi({\cal N})\right)^{(1)} \\\nonumber
&&+(-1)^{(p+1)/2}\cdot\frac{\pi}{2}\cdot  2^{p-4}
\int\left((\bar Y \wedge Z +\bar Z \wedge Y )
\wedge \chi({\cal N})\right)^{(1)}\\
&&-(-1)^{(p+1)/2}\cdot\frac{\pi}{2}\cdot2^{2(p-4)}
\int \left(\bar Z\wedge Z\wedge\chi({\cal N})\right)^{(1)}\nonumber\\
&\equiv&(-1)^{(p+1)/2}\int\left(\Delta_{BB}+\Delta_{BO^-+O^-B}+\Delta_{O^-O^-}\right)\ ,
\end{eqnarray}
where in the last line we classified the contribution to
brane-brane($BB$), brane-plane($BO$), and plane-plane($OO$) type.

Again we denote by $ch_\rho$  the trace over $\rho$
representation of $SO(2k)$. In particular,
$ch_{2k}=ch_{\overline{2k}}$ and
$ch_{2k\otimes \overline{2k}}=ch_{2k\otimes {2k}}=[ch_{2k}]^2$,
thanks to the reality of the vector representation of $SO$ groups.
Then, we find contributions with gauge group factors
\begin{equation}
\Delta_{BB} =-\frac{\pi}{2}\left( ch_{2k\otimes \overline{2k}}({\cal F})
\wedge\frac{{\cal A}({\cal T})}{{\cal A}({\cal N})}
\wedge \chi({\cal N})\right)^{(1)}_{p+1}\ ,
\end{equation}
and\footnote{A useful identity throughout here is
$$
\sqrt{{\cal A}({\cal T})}\sqrt{{\cal L}({\cal T}/4)} ={\cal A}({\cal T}/2)
$$}
\begin{eqnarray}\nonumber
\Delta_{BO^-+O^-B} &=& \frac{\pi}{2}\cdot 2^{p-4}
\left(\left[ch_{2k}({\cal F})+ch_{\overline{2k}}({\cal F}) \right]\wedge
\sqrt{\frac{{\cal A}({\cal T})}{{\cal A}({\cal N})}}\wedge
\sqrt{\frac{{\cal L}({\cal T}/4)}{{\cal L}({\cal N}/4)}}
\wedge \chi({\cal N})\right)^{(1)}_{p+1}\\\nonumber
&=&\frac{\pi}{2}\cdot2^{p-4}
\left( \left[
ch_{2k}({\cal F})+ch_{\overline{2k}}({\cal F})\right] \wedge
\frac{{\cal A}({\cal T}/2)}{{\cal A}({\cal N}/2)}\wedge
\chi({\cal N})\right)^{(1)}_{p+1}\\\nonumber
&=&\frac{\pi}{2}\left(ch_{2k}(2{\cal F}) \wedge
\frac{{\cal A}({\cal T})}{{\cal A}({\cal N})}\wedge
\chi({\cal N})\right)^{(1)}_{p+1}\ ,
\end{eqnarray}
which combine to
\begin{eqnarray}\label{D}
&&(-1)^{(p+1)/2}\left(\Delta_{BB}+\Delta_{BO^-+O^-B}\right) \nonumber\\
&=&-(-1)^{(p+1)/2}\left(\frac{\pi}{2}\left[ch_{2k\otimes 2k}(\CF)-ch_{2k}(2\CF)\right]\wedge
\frac{\CA(\CT)}{\CA(\CN)}\wedge \chi(\CN)\right)^{(1)}_{p+1}\ .
\end{eqnarray}
Purely Orientifold contribution is
\begin{eqnarray}\nonumber
(-1)^{(p+1)/2}\Delta_{O^-O^-} &=& -(-1)^{(p+1)/2}\frac{\pi}{2}\cdot2^{2(p-4)}
\left(\frac{{\cal L}({\cal T}/4)}{{\cal L}({\cal N}/4)}\wedge \chi({\cal N})\right)^{(1)}_{p+1}
\\
&=& -(-1)^{(p+1)/2}\frac{\pi}{8}\left(\frac{{\cal L}({\cal T})}{{\cal L}({\cal N})}
\wedge \chi({\cal N})\right)^{(1)}_{p+1}\ .
\end{eqnarray}
We will see later how these cancel various one-loop contributions.

Extending this to O$p^+$ plane is immediate with
\begin{equation}
S_{Op^+}=-S_{Op^-}\ ,
\end{equation}
as motivated by the fact that the two planes differ by a sign of the
charge. Again writing
\begin{eqnarray}\nonumber
\delta(S_{Op^+}+S_{Dp})
&=&(-1)^{(p+1)/2}\int \left(\Delta_{BB}+\Delta_{BO^++O^+B}+\Delta_{O^+O^+}\right)\ ,
\end{eqnarray}
the only change from O$^-$ case is  the sign flip of
$\Delta_{BO^++O^+B}=-\Delta_{BO^-+O^-B}$. As such, we have
\begin{eqnarray}\label{SP}
&&(-1)^{(p+1)/2}\left(\Delta_{BB}+\Delta_{BO^++O^+B}\right)\nonumber\\
&=&-(-1)^{(p+1)/2}\left(\frac{\pi}{2}\left[ch_{2k\otimes 2k}(\CF)+ch_{2k}(2\CF)\right]
\wedge \frac{\CA(\CT)}{\CA(\CN)}\wedge \chi(\CN)\right)^{(1)}_{p+1}\ ,
\end{eqnarray}
where the trace in $ch_\rho$ should be understood as taken in $\rho$
representations of $Sp(k)$ gauge group. The defining representation
$2k$ is pseudo-real, so the algebra goes the same as $SO(2k)$ cases.
The Orientifold contribution
\begin{eqnarray}
\Delta_{O^+O^+} =\Delta_{O^-O^-}
\end{eqnarray}
remains the same, begins quadratic in the $p$-brane charge.

Inflow in the presence of $\widetilde{\textrm{O}p^-}$'s can be similarly obtained.
Since the charge of $\widetilde{\textrm{O}p^-}$ equals to that of an $\textrm{O}p^-$
plus an half
D-brane, the obvious candidate for the CS coupling of $\widetilde{\textrm{O}p^-}$ is
\begin{equation}
S_{\widetilde{Op^-}} = \frac12\times\left( -\frac{\mu_p}{2}\int \sum_r  s^*(C_{r+1})\wedge\left[
\sqrt{\frac{{\cal A}({\cal T})}{{\cal A}({\cal N})}}
-2^{p-4}
\sqrt{\frac{{\cal L}({\cal T}/4)}{{\cal L}({\cal N}/4)}}\right] \;\right)\ .
\end{equation}
$\Delta_{BB}$ is unaffected as before, while  $\Delta_{\widetilde{O^-}B+B\widetilde{O^-}}$ is modified
as
\begin{equation}
\Delta_{B\widetilde{O^-}+\widetilde{O^-}B}=\frac{\pi}{2}\left(\left[ch_{2k}(2{\cal F})-2ch_{2k}({\cal F})\right] \wedge
\frac{{\cal A}({\cal T})}{{\cal A}({\cal N})}\wedge
\chi({\cal N})\right)^{(1)}_{p+1}\  .
\end{equation}
Thus, the analog of  (\ref{D}) and (\ref{SP}) here is
\begin{eqnarray}\label{C}
-(-1)^{(p+1)/2}\left(\frac{\pi}{2}\left[ch_{2k\otimes 2k}(\CF)-ch_{2k}(2\CF)+2ch_{2k}(\CF)\right]
\wedge \frac{\CA(\CT)}{\CA(\CN)}\wedge \chi(\CN)\right)^{(1)}_{p+1}\ .
\end{eqnarray}
Finally, the purely Orientifold contribution
may look more involved than before, but turns out to be the same:
\begin{eqnarray}\nonumber
\Delta_{\widetilde{O^-}\widetilde{O^-}} &=&
-\frac{\pi}{2}\left(\left(\sqrt{\frac{{\cal A}({\cal T})}{{\cal A}({\cal N})}}-2^{p-4}
\sqrt{\frac{{\cal L}({\cal T}/4)}{{\cal L}({\cal N}/4)}}\;\right)^2_{2p-6}\,\right)^{(1)}\wedge\chi({\cal N})_{9-p}\\\nonumber
&=& -\frac{\pi}{2}\left( \left(\frac{{\cal A}({\cal T})}{{\cal A}({\cal N})}-2^{p-3}\frac{{\cal A}({\cal T}/2)}{{\cal A}({\cal N}/2)}+2^{2(p-4)}
\frac{{\cal L}({\cal T}/4)}{{\cal L}({\cal N}/4)}\;\right)_{2p-6}\right)^{(1)}\wedge\chi({\cal N})_{9-p}\\\nonumber
&\simeq & -\frac{\pi}{2}\cdot2^{2(p-4)}\left(
\left(\frac{{\cal L}({\cal T}/4)}{{\cal L}({\cal N}/4)}\wedge \chi({\cal N})\right)_{p+3}\right)^{(1)}
\\
&=& -\frac{\pi}{8}\left(\frac{{\cal L}({\cal T})}{{\cal L}({\cal N})}
\wedge \chi({\cal N})\right)^{(1)}_{p+1} \;\;= \;\;\Delta_{O^-O^-}\ ,
\end{eqnarray}
where the equalities hold because we are supposed to
extract $p+3$-form parts of the anomaly polynomial.

\subsection{One-Loop from Open String Sector}

Consider the situation where $2k$ coincident D-branes are on the top of one of an
O$^-$, an O$^+$, or an $\widetilde{\textrm{O}^-}$ plane. There is one more type of Orientifold
plane $\widetilde{\textrm{O}^+}$, but this leads to the same gauge group as the O$^+$ case
and thus the same world-volume one-loop anomaly is induced.

First, in the presence of the O$^-$ planes, the gauge group of the open strings ending on
D$p$-branes is enhanced from $U(k)$ to $SO(2k)$. Hence a $SO(2k)$ adjoint fermion
contributes to the world-volume anomaly polynomial of amount
\begin{equation}
2\pi\cdot ch_{\frac12 2k(2k-1)}\wedge{\cal A}({\cal T})\wedge ch_{S^+}(\CN)
\end{equation}
for $4n$-dimensions, and
\begin{equation}
\pi\cdot ch_{\frac12 2k(2k-1)}\wedge{\cal A}({\cal T})\wedge [ch_{S^+}(\CN)-ch_{S^-}(\CN)]
\end{equation}
for $4n+2$-dimensions. Thanks to the reality of $SO(2k)$, two of these can be written
uniformly as
\begin{equation}
I_{1-loop}^{SO(2k)}=\pi\cdot ch_{\frac12 2k(2k-1)}\wedge{\cal A}({\cal T})\wedge [ch_{S^+}(\CN)-ch_{S^-}(\CN)]\ .
\end{equation}
By the way, we have an identity
\begin{equation}
ch_{\frac12 2k(2k\pm1)}(\CF)= \frac12 ch_{2k\otimes 2k}(\CF)\pm \frac12 ch_{2k}(2\CF)
\end{equation}
and it leads to
\begin{eqnarray}
I_{1-loop}^{SO(2k)}&=&\frac{\pi}{2}\left[ ch_{2k\otimes 2k}-ch_{2k}(2\CF)\right]\wedge{\cal A}({\cal T})\wedge [ch_{S^+}(\CN)-ch_{S^-}(\CN)]\ .
\end{eqnarray}
Again, with the identity
\begin{equation}
\frac{\chi(\CN)}{\CA(\CN)}=ch_{S^+}(\CN)-ch_{S^-}(\CN)\ ,
\end{equation}
we see that they have the precise form and the factor that can cancel inflows (\ref{D})
from $BB$ and $BO+OB$ intersection.

Similarly, the other cases follow. The symplectic case is
\begin{eqnarray}\nonumber
I_{1-loop}^{Sp(k)}&=&\pi\cdot ch_{\frac12 2k(2k+1)}\wedge{\cal A}({\cal T})\wedge [ch_{S^+}(\CN)-ch_{S^-}(\CN)]\\\nonumber
&=& \frac{\pi}{2}\,\left[ ch_{2k\otimes 2k}(\CF)+ch_{2k}(2\CF)\right]\wedge{\cal A}({\cal T})\wedge [ch_{S^+}(\CN)-ch_{S^-}(\CN)]\ ,
\end{eqnarray}
which are again canceled by the anomaly inflow $\Delta_{BB}+\Delta_{BO^++O^+B}$ (\ref{SP})
in the presence of an O$^+$ plane.
$SO(2k+1)$ type gauge theory can be also dealt with by expanding its
adjoint representation in terms of the $SO(2k)$ representation as
\begin{equation}
ch_{adj.}^{SO(2k+1)} = ch_{\frac12 2k(2k-1)+2k} = \frac12\, ch_{2k\otimes 2k}(\CF)- \frac12\, ch_{2k}(2\CF) +ch_{2k}(\CF)\ ,
\end{equation}
whereby the world-volume anomaly can be decomposed as
\begin{eqnarray}\nonumber
I_{1-loop}^{SO(2k+1)}
&=&\frac{\pi}{2}\,\left[ ch_{2k\otimes 2k}(\CF)- ch_{2k}(2\CF)+ 2ch_{2k}(\CF)\right]\nonumber\\
&&\wedge{\cal A}({\cal T})\wedge [ch_{S^+}(\CN)-ch_{S^-}(\CN)]\ ,
\end{eqnarray}
which again is neatly canceled by  $\Delta_{BB}+\Delta_{B\widetilde{O^-}+\widetilde{O^-}B}$
(\ref{C}).

Hence, we conclude that the part of anomaly and inflow that
depend on the gauge group exactly cancel regardless of the brane types,
after the overall chirality (or the orientation issue) is properly taken into account.

\subsection{On Universal Inflow $\Delta_{OO}$}\label{closed}

As $\Delta_{BB}+\Delta_{BO+OB}$ are canceled by the
open string sector one-loop, $\Delta_{OO}$ is left uncanceled so far.
Clearly this part of inflow has nothing to do with the open string degrees of freedom;
it exists even in the absence of any D-branes. As such,
$\Delta_{OO}$ should be canceled by one-loop anomaly from the closed string
spectrum. We wish to emphasize here that, even before checking
cancelation against closed string one-loop, the proposed Chern-Simons
couplings stand out because they lead to a universal inflow
\begin{equation}
\Delta_{O^-O^-}=\Delta_{O^+O^+}=\Delta_{\widetilde{O^-}\widetilde{O^-}}=
- \frac{\pi}{8}\left(\frac{{\cal L}({\cal T})}{{\cal L}({\cal N})}
\wedge \chi({\cal N})\right)^{(1)}_{p+1}\ ,
\end{equation}
from all types of Orientifold planes. This has to be the case,
as the  closed string part of the low energy spectrum
does not care what kind of projections are taken on the Chan-Paton
factors. This obvious and basic requirement is met by our
Chern-Simons couplings, which may be compared to those
in Refs.~\cite{Mukhi,Henry,JFO}.

Checking the cancelation of $\Delta_{OO}$ by closed string one-loop
for $p<9$ is a bit nontrivial, however. The simplest thing to try
would be the compact version of the same problem of $T^{9-p}/Z_2$
with $2^{9-p}$ Orientifold planes distributed, one at each fixed point.
The low energy spectra here would be identical to type I theory
compactified on $T^{9-p}$, producing one gravity multiplet and $(9-p)$
vector multiplets, transforming as vector representation under
$SO(9-p)_R$. For $p=5,7$, in particular, one can see that the
one-loop of this spectra does not completely cancel
$2^{9-p}\Delta_{OO}$.\footnote{For $p=3$, nevertheless, we do
have a complete cancelation of $\Delta_{OO}$
\begin{equation}
\label{remnant}
\int_{3+1}-2^{6}\frac{\pi}{8}\left[\frac{{\cal L}({\cal T})}{{\cal L}({\cal N})}
\wedge \chi({\cal N})\right]^{(1)}=-8\pi \int_{3+1} \chi({\cal N})^{(1)}\ ,
\end{equation}
by type I massless closed string spectra on $T^6$. The latter's
one-loop gives
\begin{eqnarray}\nonumber
&&2\pi\cdot{\cal A}({\cal T})\wedge\biggl[ ch_{S^-}({\cal N})+ [ch_{V}({\cal T})-1]\wedge ch_{S^+}({\cal N})
+ch_{S^+}({\cal N}) \wedge ch_{V}({\cal N})\biggr]\Biggl\vert_{6-form}\ .
\end{eqnarray}
Since other factors involve only $4$-forms or higher, we may replace
$ch_{S^\pm}({\cal N})$ by $ \pm \chi({\cal N})\CA(\CN)^{-1}/2$. The
Euler class $\chi(\CN)$ is 6-form, so the one-loop anomaly is
\begin{equation}
\int_{3+1}(-\pi +3\pi+6\pi)\cdot\chi({\cal N})^{(1)} =8\pi\int_{3+1} \chi({\cal N})^{(1)} \ ,
\end{equation}
canceling the inflow precisely.}
That is, unless we set the normal bundle $\CN$ to
be trivial. In the latter case, both the inflow and the one-loop
vanish individually.

In retrospect, this mismatch is to be expected since the one-loop
computation based on the massless spectra in $p+1$ dimensions only
is really computing smeared version of the anomaly, over $T^{9-p}$,
rather than the localized ones. As such, the normal bundle
information, which measures nontrivial curvature effect along $T^{9-p}$
direction to begin with, is inevitably lost along the way \cite{K3}.
One must rely on more complete information, where higher modes
such as Kaluza-Klein modes are taken into account, along the line of
Ref. \cite{ACG}. This is not an easy task, since one must also keep
track of nontrivial internal curvatures. Instead we will
consider $p=9$ case that sidesteps this complication.

\subsection{D9-O9$^-$: Green-Schwarz and Cancelation of $\Delta_{OO}$}\label{typeI}

For O9$^-$ plane,  this problem does not surface because a transverse
direction does not exist. The cancelation between $\Delta_{OO}$
and closed string one-loop is really a well-known refinement of
the type I theory Green-Schwarz mechanism, and is a standard
material (e.g. see Ref. \cite{BBS}). We record it here
for the sake of completeness.

Consider the case of $2k$ D9 and a single O9$^-$
in the coverings space. Recall that the anomaly cancelation
in type I theory involves two steps. The first is a tadpole condition
$2k=32$, leading to $SO(32)$, and the second is the Green-Schwarz mechanism
generated by the coupling of type \cite{GS},
\begin{equation}
\sim\int C_2 \wedge X_8\ ,
\end{equation}
which cancels via a modified Bianchi identity of $d(dC_2)=\cdots+ X_4$
an anomaly of type
\begin{equation}\label{Green}
\sim X_4 \wedge X_8\ .
\end{equation}
As we have
set up general anomaly inflow mechanism based on the Chern-Simons
couplings to $C$'s, we should be able to recast the Green-Schwarz
mechanism in the current, more general framework \cite{BBS}.

First of all, recall that, among the RR tensor fields of type IIB
theory, only $C_2$ and its dual $C_6$ survive the Orientifold
projection to type I theory. The Chern-Simons coupling for $p=9$ is
\begin{eqnarray}\label{D9}
S_{CS} &= &\frac12\cdot\frac{\mu_9}{2} \int  \left({\sum_r}'  C_{r+1}\wedge [Y({\cal F},{\cal R})-32Z({\cal R})]\right) \\
&=&\frac12\cdot\frac{\mu_9}{2} \int  C_{2}\wedge [Y({\cal F},{\cal R})-32Z({\cal R})]_8
+C_{6}\wedge [Y({\cal F},{\cal R})-32Z({\cal R})]_4\ ,\nonumber
\end{eqnarray}
where $Y$ and $Z$ are defined as (\ref{definition}).  $C_0, C_4, C_8$ are projected out
while $C_{10}$ does not exist as a dynamical field, so that anomaly inflow
has the polynomial,
\begin{equation}
-\frac{\pi}{2}\,\Biggl([Y({\cal F},{\cal R})-32Z({\cal R})]_4+ [Y({\cal F},{\cal R})-32Z({\cal R})]_8\Biggr)^2\ .
\end{equation}
When expanded, the inflow can be also organized as
\begin{eqnarray}\label{9}
\Delta'=-\frac{\pi}{2}\left[ch_{2k\otimes 2k}({\cal F})\wedge{\cal A}({\cal R})\right]^{'(1)}_{10}
+ \frac{\pi}{2}\left[ch_{2k}(2{\cal F})\wedge {\cal A}({\cal R})\right]^{'(1)}_{10}
-\frac{\pi}{8} [{\cal L}({\cal R})]^{'(1)}_{10}\ ,
\end{eqnarray}
where each terms are from $BB$, $BO+OB$ and $OO$ intersections respectively,
and the prime $'$ signifies that we dropped terms proportional to
$Y_0$ and $Z_0$ when expanding $(Y-32Z)^2$ to compute $\Delta$.

On the other hand, from the supergravity multiplet, we have
a left-handed gravitino and a right-handed dilatino which are both
Majorana-Weyl. They carry 12-form anomaly polyniomial of amount
\begin{equation}
I_{closed}=\frac{2\pi}{2}\left[I_{3/2}(R)-I_{1/2}(R)\right]_{12}
=\frac{\pi}{8}\cdot {\cal L}({\cal R})_{12}\ .
\end{equation}
In the open string sector,
there are Majorana-Weyl gauginos in the adjoint representation of $SO(2k)$.
They contribute 12-form of
\begin{eqnarray}\nonumber
I_{open}&=&\frac{2\pi}{2}\cdot\left[{\cal A}({\cal R})
\wedge ch_{\frac{2k(2k-1)}{2}}({\cal F})\right]_{12}\\
&=& \frac{\pi}{2}\cdot \left[{\cal A}({\cal R})\wedge ch_{2k\otimes 2k}({\cal F})-
 {\cal A}({\cal R})\wedge ch_{(2k)}(2{\cal F})\right]_{12}\ .
\end{eqnarray}
These two can be combined,
\begin{equation}\label{total}
I_{closed}+I_{open}= \frac{\pi}{2}
\Biggl[\biggl(Y(\CF,\CR)-32Z(\CR)\biggr)^2\Biggr]_{12}\ ,
\end{equation}
which superficially looks similar to the inflow up to sign.

Note that this one-loop anomaly does not match the inflow
above. The inflow cancels
\begin{eqnarray}
\pi\,[Y({\cal F},{\cal R})-32Z({\cal R})]_4\wedge
[Y({\cal F},{\cal R})-32Z({\cal R})]_{8}\ ,
\end{eqnarray}
but the other terms in (\ref{total}) remain. Of these,
$([Y({\cal F},{\cal R})-32Z({\cal R})]_6)^2$
piece vanishes identically on its own, so the discrepancy is
\begin{eqnarray}
\pi\,[Y({\cal F},{\cal R})-32Z({\cal R})]_0
\wedge [Y({\cal F},{\cal R})-32Z({\cal R})]_{12}\ ,
\end{eqnarray}
bringing us to the usual tadpole condition of type I theory
\begin{equation}
Y_0({\cal F},{\cal R})-32Z_0({\cal R}) = 2k-32 =0\ ,
\end{equation}
for a complete anomaly cancelation.

With this tadpole condition obeyed, the closed string sector
one-loop $I_{closed}$  cancels on its own against the purely
Orientifold contribution,
\begin{equation}
\Delta'_{OO}=-(\pi/8)[\CL(\CR)]^{'(1)}_{10}
=-2^9\pi[\CL(\CR)/4]^{'(1)}_{10}
= - 2^9\pi \bigl[Z(\CR)^2]^{'(1)}_{10}\ ,
\end{equation}
and the open string one-loop $I_{open}$
cancels the rest of the inflow, $\Delta'-\Delta_{OO}'$,
coming from $BB$ and $BO+OB$ intersections.

\section{Summary}

In this note, we re-examined in detail the inflow mechanism
onto the various world-volume theories, with the aim to clear
up loose ends on I-brane/D-brane inflow.

We started with a review of the M5-brane inflow and the
I-brane/D-brane inflow. Historically, the M5-brane anomaly
inflow was studied in two steps. Some anomaly inflow onto the M5
arises rather straightforwardly out of a Chern-Simons
coupling between antisymmetric tensor field $C_3$ and
spacetime curvature, but this turned out inadequate for
the $SO(5)_R$ axial anomaly, as was first
noted by Witten (\ref{leftover}). For the I-brane/D-brane
anomaly, the necessary topological coupling lives on the
world-volume, rather than on spacetime, and is  linear
in the RR antisymmetric tensor fields. The
difference from the M5 example is that many RR fields
enter the inflow mechanism simultaneously. For the latter also,
the cancelation against one-loop was partial in that there
were, so-called self-dual cases, such as D3-branes, where
the necessary inflow were apparently  absent.

The problem of the axial anomaly deficit for the M5-brane was
eventually
solved by FHMM, who noted that one should carefully treat the
singularity at the position of the M5-brane. They replaced
the usual naive delta-function by a covariantized and smeared
version, and at the same time demanded a regularity of the
resulting field strengths. It requires a particular transformation
rule for the three-form gauge field,  $C_3$, and as a consequence,
they yield a right inflow from Chern-Simon terms of type, $C_3\wedge dC_3\wedge dC_3$.

It is then almost immediate that the regularity requirement on the field
strengths should be also obeyed in the I-brane/D-brane inflow
mechanism. With the Thom class for IIB D-branes,
$\tau_n=d[\rho \cdot\hat e_{n-1}]=d[\rho \cdot(d\psi_{n-2}+\Omega_{n-1})]$,
we should choose its descent as
\begin{equation}
\tau^{(0)}_{n-1} = -d\rho\wedge \psi_{n-2} +\rho\cdot\Omega_{n-1}\ ,
\end{equation}
when solving the Bianchi identity.
Note that this choice of $\tau^{(0)}_{n-1}$ has a nontrivial transformation
under the  normal bundle gauge transformation, unlike the naive, invariant,
but singular choice $\rho\cdot \hat e_{n-1}$.
This modifies the variation of the RR tensor fields (\ref{new}),
and results in a new form of inflow  (\ref{variation}). For $p\ge 5$,
the new inflow agrees with the old one, up to local counter-terms
on world-volume, while for $p=3$ the inflow no longer vanishes
and neatly cancels the one-loop anomaly of the D3-brane open string sector.

Furthermore, the modified inflow mechanism is such that
the correct answer emerges regardless of specific form of
the Chern-Simons couplings used, be it $S_{CS}$ (\ref{CS2}) or
$S_{CS}'$ (\ref{CS3}),
\begin{equation}
\delta S_{CS}\simeq \delta S_{CS}'\ ,
\end{equation}
where the  difference between the two
amounts to a local counter-term on the world-volume.
This should be contrasted to the previous inflow
mechanism that resulted in correct answer only from
$ S_{CS}'$. With old $\tilde\delta$ of (\ref{delta'}),
$\tilde\delta S_{CS}'$ is NOT equivalent
to $\tilde\delta S_{CS}$ up to a local counter-term,
even when it produces correct inflow. This was a little
curious and potentially confusing, all the more, as
$S_{CS}$ looks by far more natural than $S_{CS}'$ yet
failed to generate the anticipated inflow via $\tilde\delta$.
By the physically
motivating revised  transformation rule $\delta$ (\ref{new}), instead,
we also cured this  phenomenon.

Finally, we extended this to the theories including the Orientifold
planes. Curiously, there appears to be no complete consensus on the
gravitational Chern-Simons couplings on some of Orientifold planes.
We settled this by requiring cancelation between anomaly inflow and
one-loop contributions from the open string sector and the closed
string sector. We computed the most general one-loop anomaly of the maximally
supersymmetric Yang-Mills theories in all even dimensions, and determined
the necessary Chern-Simons couplings on the four types of Orentifold
planes as in (\ref{Ominus}), (\ref{Otilde}) and (\ref{Oplus}).
Gauge theory one-loop anomaly completely cancels out part of the
inflow that involves D-branes. The pure Orientifold part of
inflow, $\Delta_{OO}$, to be canceled by the closed string sector
one-loop,  is also shown to be universal, i.e., independent of types
of the Orientifold, which, we argued, is by itself a nontrivial
consistency check. Our result for O$^+$ planes, in particular,
agrees with Refs.~\cite{DFM}\cite{Scrucca:1999jq}.

\vskip 1cm
\centerline{\bf\large Acknowledgments}
\vskip 5mm\noindent
We are grateful to M{\aa}ns Henningson for explaining his work
on dyonic strings, and other discussions at early stage of this
work, and (P.Y.) thank organizers and staffs of Simons Summer
Workshop in Mathematics and Physics 2011 for hospitality.
This work is supported by the National Research Foundation of Korea
(NRF) funded by the Ministry of Education, Science and Technology
with grant number 2010-0013526 (P.Y.) and 2009-0076297 (H.K.), and
by the project of Global Ph.D. Fellowship which
National Research Foundation of Korea conducts from 2011 (H.K.).

\vskip 1cm

\appendix

\section{$1/2$  in the Minimal Couplings}

In this appendix, we show the simplest example of duality symmetric
formulation of $p$-form theory and illustrate how the additional
factor $1/2$ in the minimal coupling is necessary.
For more comprehensive studies, see Refs.~\cite{SS}\cite{Deser}.
The simplest example is a Maxwell theory in $d=4$, whose
electric form is
\begin{equation}\label{maxwell}
\frac12\int dx^4 \left(\bE\cdot\bE-\bB\cdot \bB\right) - e\int dt \left(\bA\cdot \dot{\bf r}+\phi\right)\ ,
\end{equation}
with $A_0=\phi$ and electric charge $e$. Coupling this to both electric and
magnetic charges simultaneously is more involved but can
be done relatively easily at the expense of explicit Lorentz invariance.
For this, one
introduces two sets of Maxwell fields,
\begin{eqnarray}\label{B}
\bB^{(a)}=\NA\times \bA^{(a)} +\cdots \ , 
\qquad \bE^{(a)}=\partial_t \bA^{(a)}-\NA \phi^{(a)} +\cdots \ ,
\end{eqnarray}
for $a=1,2$, where the ellipses encode possible violations of the Bianchi
identities,
and invents a duality symmetric action,
\begin{eqnarray}\label{dual}
&&\frac12\int dx^4 \left(\bB^{(1)}\cdot \bE^{(2)} -\bB^{(2)}\cdot \bE^{(1)}
-\bB^{(1)}\cdot \bB^{(1)}-\bB^{(2)}\cdot \bB^{(2)}  \right) \nonumber\\
&& -\frac 12 \; q^{(2)}\int dt\left(\bA^{(1)}\cdot \dot{\bf r}+\phi^{(1)}\right)
+\frac 12 \; q^{(1)}\int dt \left(\bA^{(2)}\cdot \dot{\bf r}+\phi^{(2)}\right)\ .
\end{eqnarray}
The Gauss constraints are
\begin{equation}\label{Gauss}
\NA\cdot \bB^{(a)}+q^{(a)} \delta({\bf x}-{\bf r}(t))=0\ ,
\end{equation}
which also imply
\begin{equation}\label{BB}
\partial_t \bB^{(a)}-\NA\times \bE^{(a)} - q^{(a)}\dot{\bf r}\, \delta({\bf x}-{\bf r}(t))=0\ .
\end{equation}
Note that $q^{(1)}$ and $q^{(2)}$ act like magnetic charges of
$-\bB^{(1)}$ and $-\bB^{(2)}$, although they will be eventually
identified as  electric charges of $-\bE^{(2)}$ and $\bE^{(1)}$ later.

Consider the simple case of  $(q^{(1)},q^{(2)})=(0,e)$. With $q^{(1)}=0$, equation
of motion for $\bA^{(2)}$ 
is solved as
\begin{equation}\label{A2}
\bE^{(1)}+\bB^{(2)}=0\ ,
\end{equation}
killing off $-\bB^{(2)}\cdot \bE^{(1)} -\bB^{(2)}\cdot \bB^{(2)}$ in the action,
and one of the Gauss constraints (\ref{Gauss}) becomes more conventional,
\begin{equation}
\NA\cdot \bE^{(1)}
=e\,\delta({\bf x}-{\bf r}(t)) \ .
\end{equation}
The remaining terms in the action are
\begin{eqnarray}
&&\frac12\int dx^4 \left(\bB^{(1)}\cdot \bE^{(2)}
-\bB^{(1)}\cdot \bB^{(1)} \right)
- \frac 12 \,e\int dt\,\left(\bA^{(1)}\cdot \dot{\bf r}+\phi^{(1)}\right)\nonumber \\
&=&\frac12\int dx^4 \left(\bA^{(1)}\cdot \NA\times \bE^{(2)}
-\bB^{(1)}\cdot \bB^{(1)} \right)
- \frac 12 \,e\int dt\,\left(\bA^{(1)}\cdot \dot{\bf r}+\phi^{(1)}\right)\nonumber \\
&=&\frac12\int dx^4 \left(-\bA^{(1)}\cdot \partial_t \bE^{(1)}
-\bB^{(1)}\cdot \bB^{(1)} \right)
- \frac 12 \,e\int dt\,\left(2\bA^{(1)}\cdot \dot{\bf r}+\phi^{(1)}\right)\ ,
\end{eqnarray}
where we used (\ref{BB}) and (\ref{A2}). With $q^{(1)}=0$,
the Bianchi identity of the first gauge field holds, so
we may use $\partial_t\bA^{(1)}=\bE^{(1)}+\NA \phi^{(1)}$.
Integrating by parts and using the
Gauss constraint (\ref{Gauss}) again,
\begin{eqnarray}
&&\frac12\int dx^4 \left(\bE^{(1)}\cdot \bE^{(1)}
-\bB^{(1)}\cdot \bB^{(1)} \right)
- \,e\int dt\,\left(\bA^{(1)}\cdot \dot{\bf r}+\phi^{(1)}\right)\ ,
\end{eqnarray}
we find the usual Maxwell action with electric charge $e$,
without the factor $1/2$. This shows that the correct equation
of motion emerges, even though the minimal coupling has an
unfamiliar factor $1/2$.

One may repeat the exercise, with a static charge $(q^{(1)},q^{(2)})=(-g,0)$
instead, by integrating out the other Maxwell fields. The equation
of motion from $\bA^{(1)}$ is solved as
\begin{equation}
\bE^{(2)}-\bB^{(1)}=0\ , 
\end{equation}
and, a similar procedure produces
\begin{eqnarray}
&&\frac12\int dx^4 \left(\bE^{(2)}\cdot \bE^{(2)}
-\bB^{(2)}\cdot \bB^{(2)} \right)
-  g \int dt\,\left(\bA^{(2)}\cdot \dot{\bf r}+\phi^{(2)}\right)\ ,
\end{eqnarray}
again without the factor $1/2$.

The factor $1/2$ in the symmetric formulation is also consistent
with the Dirac quantization. Suppose that
a particle with $(q^{(1)},q^{(2)})=(0,e)$ is present in the vicinity of
another with $(q^{(1)}, q^{(2)}) =(-g,0)$. In usual electric formulation,
where $e$ and $g$ are electric and magnetic charges of $\bA$,
the quantization comes from the invisibility of the Dirac string
of the latter as  the former circles it along a small loop $\gamma$.
The phase shift on the former wavefunction would be
\begin{equation}
 e\int_\gamma  \bA_g\cdot d{\bf x}=4\pi \,e\cdot g\ .
\end{equation}
In the duality symmetric formulation, however, both particles
generate Dirac strings, due to (\ref{Gauss}), and
we have two such contributions to the
phase shift. For simplicity, we may imagine the two Dirac
strings stretched along positive and negative $z$-axis, respectively.
As the first particle moves along $\gamma$, encircling
a Dirac string, the second particle also circles
around a Dirac string of the other, along $-\gamma$, once.
The combined phase shift is
\begin{equation}
\frac12\, e\int_\gamma  \bA^{(1)}_g\cdot d{\bf x}
+\frac12 \,g\int_{-\gamma} \bA^{(2)}_{-e}\cdot d{\bf x}
=2\pi \,e\cdot g+2\pi g\cdot e=4\pi \,e\cdot g\ .
\end{equation}
Thus, the factor $1/2$ is not only consistent with but
necessary for preserving the usual Dirac quantization condition.

\section{Duality Symmetric Action for RR Fields}

What we saw in the previous section extends to the collection of
RR tensors in type II theories as
\begin{eqnarray}\label{SDII}
&&\frac{1}{4\kappa_{10}^2}\int d^{10} x \sum_n\left[~ B_{n}\wedge
*(B_{n})-(-1)^{(n+\epsilon-2)/2}B_{n}\wedge E_{10-n}
\right]\nonumber \\
&& +\frac{1}{2}\sum_p\mu_{p} 
\int_{Dp} \sum_q s^*(C_{q+1})\wedge Y_{p-q}\ ,
\end{eqnarray}
with the universal  $1/2$ factor in the minimal coupling.
The Hodge star operation is taken to act on the right,
\begin{equation}
*B_{i_{n+1}\cdots i_{d}}=\frac{1}{n!}\,B_{i_1\cdots i_n}
\epsilon^{i_1\cdots i_n}_{\;\;\;\;\;\;\;\;\;\;\; i_{n+1}\cdots i_d}\ ,
\end{equation}
 so that the
first term is negative definite with $(-++\cdots+)$ signature.\footnote{This
type of duality symmetric action for tensor fields works in $d=4k+2$
with (anti-)self-dual middle form, while, for $d=4k$, the middle form must be doubled
as in Appendix A and the sign for half of kinetic terms must be flipped.}
As in Appendix A, $H_n=dC_{n-1}$ is
split into space-like $B$ and time-like $E$.  More precisely,
if we split $C={\bf C}+\Phi$
with the time-like part $\Phi$ and similarly $d={\bf d}+d_t$,
we have
\begin{equation}
B={\bf d}{\bf C}+\cdots \ ,\qquad E= d_t{\bf C}+{\bf d}\Phi+\cdots \ ,
\end{equation}
again up to terms in the ellipses that violate the naive Bianchi identity.

Note that, using the same line of argument
as in Appendix A, we obtain
$*H_n=(-1)^{(n-2+\epsilon)/2}H_{10-n}$, and thus the
first line is the duality symmetric kinetic term that we implicitly
used in this note.
The field equation and the Bianchi identity are
\begin{equation}
d(*(H_{r+2}))=-(-1)^r2\kappa_{10}^2\mu_{r}\tau_{9-r} \ , \qquad dH_{r+2}= (-1)^{(r+\epsilon)/2}2\kappa_{10}^2\mu_{6-r}\tau_{r+3} \ ,
\end{equation}
with a single D$r$-brane and with a single D$(6-r)$-brane, respectively, and
all curvatures turned off.
More generally, the Bianchi identity with curvatures turned on and
other D-branes present is
\begin{equation}
dH_{r+2}=\sum_p (-1)^{(r+\epsilon)/2} 2\kappa_{10}^2\mu_{p}\, Y_{p+r-6} \wedge\tau_{9-p}
=-\sum_p (-1)^{(-p+\epsilon)}2\kappa_{10}^2\mu_{p}\, \bar Y_{p+r-6} \wedge\tau_{9-p}  \ ,
\end{equation}
where the right hand side represents induced D$(6-r)$-brane charges
on D$p$-branes.

For $4$-form field $C_4$, with its dual also a 4-form, we need
to be  more careful. $H_5$ is
constrained to be self-dual, so degrees of freedom
counting suggests only one $C_4$ enter the action. Indeed,
it is known \cite{Deser}  that  such a kinetic term for a single $C_4$
generates self-duality constraint from equation of motion. Thus, the pertinent
question is whether the single minimal coupling of D3 to $C_4$ is also
consistent with self-duality of $H_5$ and whether the same factor of $1/2$
in the minimal coupling leads to correctly Dirac-quantized sources.
Here, we will show that the homogeneous and the inhomogeneous
part of $H_5$ are respectively self-dual, as a consequence of the
above kinetic term.

First, we review the source-free case for $n=5$ and $\epsilon=1$.
With the spatial indices
denoted by capital roman characters, $A$, $B$, etc, note that
\begin{equation}
H_{ABCDE} = \partial_AC_{BCDE}+\cdots+ \partial_EC_{ABCD}\ ,
\end{equation}
where the sum is over the cyclic permutations.
In the absence of source, the action (\ref{SDII}) reduces
\begin{equation}\label{sourcefree}
I=\frac12\int d^{10}x \left[E^{ABCD}B_{ABCD}-B^{ABCD}B_{ABCD}\right]\ ,
\end{equation}
where we define $B$ and $E$ by the magnetic and electric components of $H$,
\begin{eqnarray}
B_{ABCD} &=& \frac{1}{120} \epsilon^{ABCDEFGHI}H_{EFGHI}\ ,\\
E_{ABCD} &=& -H^{0ABCD}\ ,
\end{eqnarray}
with the indices $A\sim I=1\cdots 9$.
Then variation of (\ref{sourcefree}) with $C_{ABCD}$ gives
\begin{equation}
\frac{1}{2}\left[\frac{1}{12}\epsilon^{FGHIEABCD}\partial_E\left(E^{FGHI}-B^{FGHI}\right)\right]=0\ ,
\end{equation}
and we can choose $C_{0ABC}$ so that the solution can be written as
\begin{equation}\label{Selfdual}
E^{FGHI}=B^{FGHI}\ ,
\end{equation}
which is equivalent to $H_5=*H_5$, the self-duality equation.

The relation (\ref{Selfdual}) also holds in the presence of a source term.
Now the action is written in a form
\begin{eqnarray}\nonumber
I=&&\frac12\int d^{10}x \left[E^{ABCD}B_{ABCD}-B^{ABCD}B_{ABCD}\right]\\
&&+ \frac12\int d^{10}x \left[C_{ABCD}J^{ABCD}+C_{0ABC}J^{0ABC}\right]\ ,
\end{eqnarray}
where $J^{ABCD}$ is a current source.
We also add possible contributions of source to the field strengths by
\begin{eqnarray}\label{B_ABCD}
B_{ABCD} &=& \frac{1}{120} \epsilon^{ABCDEFGHI}H_{EFGHI}-G^{0ABCD}\ ,\\\label{E_ABCD}
E_{ABCD} &=& -H^{0ABCD}+F^{ABCD}\ .
\end{eqnarray}
Here, the additional term $G$ can be thought of as the inhomogeneous solution,
\begin{equation}\label{current}
\partial_E G^{ABCDE}-\partial_0 G^{0ABCD} = -J^{ABCD}\ ,
\end{equation}
which is consistent with the Bianchi identity
\begin{equation}\label{hahaha}
\partial_D B^{ABCD} = -\partial_D G^{0ABCD} = -J^{0ABC}\ .
\end{equation}
Finally, variation of the action with $C_{ABCD}$ gives an equation of motion,
\begin{eqnarray}\nonumber
&&\frac{1}{12}\epsilon^{FGHIEABCD}\partial_E\left(B_{FGHI}-E_{FGHI}\right)\\
&&-\partial_0G^{0ABCD}+\frac{1}{24}\epsilon^{ABCDEFGHI}\partial_E F_{FGHI}+J^{ABCD}=0\ .
\end{eqnarray}
Thanks to the self-duality relation in the absence of source and (\ref{current}),
we can conclude from the equation of motion that $F^{ABCD}$ should satisfy
\begin{equation}
F^{ABCD} = \frac{1}{120}\epsilon^{ABCDEFGHI}G_{EFGHI}\ .
\end{equation}
Note that this equation implies the source term contributions, $G$ and $F$ of
(\ref{B_ABCD}) and (\ref{E_ABCD}), also should be self-dual, which requires the
dyonic source of the equal magnetic and electric charge.
Then again by a suitable choice of $C_{0ABC}$, we can see that the self-duality equation,
\begin{equation}\label{B=E}
B^{ABCD}=E^{ABCD}\ ,
\end{equation}
holds in general, even in the presence of the self-dual dyonic sources.
Finally, combining (\ref{hahaha}) and (\ref{B=E}), the correct field equations
for $C_4$ with a dyonic source are induced, justifying the minimal coupling to
$C_4$ only with the by-now-familiar factor $1/2$.

One consistency check is again
the Dirac quantization condition. When one D3 revolves around another's
Dirac-string-like singularity,
we find the phase picked up in the process is
\begin{equation}
2\times\frac{\mu_3}{2}\times 2\kappa_{10}^2\mu_3 =2\kappa_{10}^2\mu_3^2=2\pi \ ,
\end{equation}
where the overall factor 2 occurs because each D3 acts as a magnetic source
for the other's electric charge. This again makes the Dirac-string-like singularities
 invisible. This is not much different from
other dual pairs, where $2\pi$ is achieved as
\begin{equation}
\frac{\mu_p}{2}\times 2\kappa_{10}^2\mu_{6-p}+\frac{\mu_{6-p}}{2}\times 2\kappa_{10}^2\mu_p =2\kappa_{10}^2\mu_p\mu_{6-p}=2\pi \ ,
\end{equation}
instead, for $p\neq 3$, again in the duality symmetric formulation.

\section{Characteristic Classes: Brief Summary}

We list characteristic classes that appear in the anomaly inflow
consideration. The Chern class is
\begin{equation}
ch_{\mathbf R}(\CF)\equiv \tr_{\mathbf R} e^{\CF/2\pi}= \sum_i e^{x_i}\ ,
\end{equation}
where ${\mathbf R}$ denotes the relevant representation,
and $x_i$ are the two-form-valued eigenvalues of
$\CF/2\pi$ in the representation ${\mathbf R}$. The A-roof genus and the
Hirzbruch class for
$SO$ bundle are, in terms of skew-eigenvalue 2-forms
$y_i$ of $R/2\pi$,
\begin{equation}
\CA (R)\equiv \prod_i \frac{y_i/2}{\sinh(y_i/2)}\ ,\qquad
\CL (R)\equiv \prod_i \frac{y_i}{\tanh(y_i)}\ .\qquad
\end{equation}
These can also be expanded in term of Pontryagin classes,
\begin{equation}
p_1(R)=\sum_i y_i^2\ , \quad p_2(R)=\sum_{i<k}y_i^2 y_k^2\ , \quad p_3(R)=\sum_{i<k<l}y_i^2 y_k^2 y_l^2\ ,
\end{equation}
and so on. Finally, the Euler class is
\begin{equation}
\chi(R)=\prod_i y_i\ .
\end{equation}
With these, we see
\begin{equation}
\frac{\chi(R)}{\CA(R)} =\prod_i \frac{\sinh(y_i/2)}{y_i/2}\prod_j y_j
= \prod_i \left(e^{y_i/2}-e^{-y_i/2}\right) = ch_{S^+}(R)-ch_{S^-}(R)\ ,
\end{equation}
for example, giving us  the central identities in relating the inflow to the one-loop
contribution, and also
\begin{equation}
\CA (R)\CL(R/4)=\prod_i \frac{2(y_i/4)^2}{\sinh(y_i/2)\tanh(y_i/4)}= \prod_i \frac{(y_i/4)^2}{\sinh(y_i/4)^2}=\CA(R/2)^2\ ,
\end{equation}
which was useful in section 5.

\end{document}